\documentclass[twocolumn,prb,showpacs,preprintnumbers,amsmath,amssymb]{revtex4}

\usepackage{graphicx}
\usepackage{dcolumn}
\usepackage{bm}

\begin{document}

\title{Construction of microscopic model for ${\bm f}$-electron systems \\
on the basis of ${\bm j}$-${\bm j}$ coupling scheme}

\author{Takashi Hotta$^1$ and Kazuo Ueda$^{2,1}$}

\affiliation{
$^1$Advanced Science Research Center,
Japan Atomic Energy Research Institute,
Tokai, Ibaraki 319-1195, Japan \\
$^2$Institute for Solid State Physics,
University of Tokyo, Kashiwa, Chiba 277-8581, Japan
}

\date{\today}

\begin{abstract}
We construct a microscopic model for $f$-electron systems, composed of
$f$-electron hopping, Coulomb interaction, and crystalline electric field
(CEF) terms.
In order to clarify the meaning of one $f$-electron state, here
the $j$-$j$ coupling scheme is considered, since the spin-orbit interaction
is generally large in $f$-electron systems.
Thus, the $f$-electron state at each site is labelled by $\mu$, namely,
the $z$-component of total angular momentum $j$.
By paying due attention to $f$-orbital symmetry,
the hopping amplitudes between $f$-electron states
are expressed using Slater's integrals.
The Coulomb interaction terms among the $\mu$-states are written by
Slater-Condon or Racah parameters.
Finally, the CEF terms are obtained from the table of Hutchings.
The constructed Hamiltonian is regarded as an orbital degenerate Hubbard model,
since it includes two pseudo-spin and three pseudo-orbital degrees of freedom.
For practical purposes, it is further simplified into a couple of
two-orbital models by discarding one of the three orbitals.
One of those simplified models is here analyzed using the exact
diagonalization method to clarify ground-state properties
by evaluating several kinds of correlation functions.
Especially, the superconducting pair correlation function
in orbital degenerate systems is carefully calculated based on the concept
of off-diagonal long-range order.
We attempt to discuss a possible relation of the present results 
with experimental observations for recently discovered
heavy fermion superconductors CeMIn$_5$ (M=Ir, Co, and Rh),
and a comprehensive scenario to understand
superconducting and antiferromagnetic tendencies in the so-called
``115'' materials such as CeMIn$_5$, UMGa$_5$, and PuCoGa$_5$ from
the microscopic viewpoint.
\end{abstract}

\pacs{71.27.+a, 71.10.-w, 71.70.Ej, 74.25.-q}

\maketitle

%
%
\section{Introduction}

Since the pioneering discovery of superconductivity in CeCu$_2$Si$_2$,
\cite{Steglich}
elucidation of the mechanism of unconventional superconductivity
in heavy fermion compounds has been one of the central issues
both in the experimental and theoretical research fields
of condensed matter physics.
In particular, further discoveries of superconductivity in uranium compounds
such as UBe$_{13}$,\cite{UBe13} URu$_2$Si$_2$,\cite{URu2Si2}
UPt$_3$,\cite{UPt3} UPd$_2$Al$_3$,\cite{UPd2Al3} and
UNi$_2$Al$_3$\cite{UNi2Al3}
have triggered a rapid increase of vigorous investigations
on exotic properties of superconductivity in $f$-electron materials.

Recently there has occurred a second rush of new discoveries
of superconductivity in $f$-electron compounds.
One such is the coexistence of superconductivity and ferromagnetism
in UGe$_2$\cite{UGe2} and URhGe.\cite{URhGe}
After this discovery, much attention has been attracted to the mechanism
of unconventional triplet superconductivity in $f$-electron systems,
since it is naively believed that triplet pairing can coexist
in the ferromagnetic phase.
A second example is the family of Ce-based heavy fermion superconductors
CeMIn$_5$ (M=Ir, Rh, and Co).\cite{Ce115-1,Ce115-2,Ce115-3}
Surprisingly, CeCoIn$_5$ exhibits a superconducting transition temperature
$T_{\rm c}$=2.3K, which is the highest among those yet observed for
heavy fermion materials at ambient pressure.
On the other hand, CeIrIn$_5$ shows $T_{\rm c}$=0.4K,
which is much less than that of CeCoIn$_5$.
Note that CeRhIn$_5$ is an antiferromagnet with a N\'eel temperature
$T_{\rm N}$=3.8K at ambient pressure, while under high-pressure
it becomes superconducting at $T_{\rm c}$=2.1K.
It is interesting that the ground state properties of these
compounds are easily changed by transition metal ions.
Third is the superconductivity of the filled skutterudite compound
PrOs$_4$Sb$_{12}$ with $T_{\rm c}$=1.85K and
a mass enhancement factor as large as fifty.\cite{PrOs4Sb12}
Again, a relatively high $T_{\rm c}$ is surprising, while an interesting
point is that the ground state is suggested to belong to $\Gamma_3$,
leading to a potential role played by quadrupole fluctuations
in the mechanism of superconductivity.
Another amazing result is the discovery of superconductivity in PuCoGa$_5$
with $T_{\rm c}$=18.5K.\cite{PuCoGa5}
Although it is difficult to synthesize this compound and to measure various
physical quantities, owing to the inclusion of plutonium,
this is truly high $T_{\rm c}$ and its mechanism is rather mysterious.

Among several remarkable features both of new and old $f$-electron
superconductors, one interesting and important point is
the common observation of anisotropic Cooper-pairing,
experimentally suggested by the power-law behavior of physical quantities
in the low-temperature region.
As is well known, the value of the power sensitively depends on the node
structure of the gap function on the Fermi surface.
Thus, by analysing carefully the temperature dependence of physical quantities
in experiments, it is possible to deduce the symmetry of the Cooper pairs
under the group-theoretical restrictions.
As a result of this phenomenological analysis, it has been clarified that
the superconducting gap in $f$-electron materials has nodes
on the Fermi surface, indicating that non $s$-wave Cooper pairing occurs
in heavy fermion superconductors.

The appearance of non $s$-wave pairing itself is easily understood
as an effect of strong short-range correlation,
if we recall the fact that the Cooper pair is composed of a pair of
$f$ electrons with heavy effective mass.
Since superconductivity in heavy fermion materials is a second-order transition
phenomenon, the Ginzburg-Landau theory is still applicable to $f$-electron
superconductors if we take into account
the symmetry of the Cooper pairs as well as the crystal structure
with the help of group-theoretical arguments.
Such efforts have been quite useful to promote the understanding of
the nature of unconventional superconductivity without considering the
mechanism of the Cooper-pair formation.\cite{Sigrist}

As for the microscopic aspects of the mechanism of unconventional
superconductivity, it has at least been confirmed that the strong short-range
Coulomb interaction plays a crucial role.
For instance, it brings about antiferromagnetic spin fluctuations,
\cite{Miyake} which induce singlet $d$-wave superconductivity,
\cite{Moriya} consistent with the node structure suggested by some
experiments.
It is true that in some $f$-electron materials, anisotropic singlet pairing is
stabilized, but in UPt$_3$, it has been experimentally confirmed that
triplet pairing occurs.\cite{Tou}
Quite recently, also in UNi$_2$Al$_3$, a possibility of triplet pairing
has been discussed.\cite{Ishida1}
Unfortunately, the triplet pair formation as well as the coexistence of
superconductivity and ferromagnetism cannot be simply explained
as the effect of antiferromagnetic spin fluctuations.

One may naively consider that instead, we can include the effect
of ferromagnetic spin fluctuations, which was proposed for the origin
of superfluidity in helium 3.
However, we then immediately face another problem, i.e.,
how to obtain such
fluctuations from the strong short-range Coulomb interaction.
It may be possible to derive this based on a realistic electronic
band structure, but we do not find that paramagnons are always dominant
in the spin fluctuation spectrum when we survey the magnetic
properties of those materials.
In fact, in ruthenate Sr$_2$RuO$_4$,\cite{Maeno}
which is experimentally confirmed to be a triplet superconductor,\cite{Ishida2}
instead we find that the incommensurate antiferromagnetic spin fluctuation
is enhanced, as observed in neutron scattering experiments.\cite{Sidis}
Namely, in contrast to our naive expectation, paramagnons do not
seem to play a central role in the occurrence of triplet superconductivity
in the solid state.
Thus, it is still a puzzling and challenging problem to clarify
what is the key issue which determines the symmetry of Cooper pairs,
singlet or triplet, in $f$-electron superconductors.

In order to understand this point, it is necessary to perform a microscopic
analysis based on a model Hamiltonian appropriate for $f$-electron systems,
but unfortunately, we find such a model to be lacking.
It may be possible to consider this problem based
on the single-band Hubbard model,
which is believed to describe well the properties of
strongly correlated electron systems.
In fact, several basic features of high $T_{\rm c}$ cuprates have been
understood based on the Hubbard model by analyzing it using various kinds
of analytical and numerical techniques.\cite{Dagotto1}
However, this model is too simplified to include differences among
materials characterized by $orbital$ degrees of freedom,
although recently, the active role of orbitals has been widely recognized
in several materials as a means for understanding novel magnetism and
unconventional superconductivity.
For instance, in manganites, $e_{\rm g}$ orbitals play a primary role
for electronic properties,\cite{Dagotto2}
since orbital ordering has been successfully
observed in manganites by means of the resonant X-ray scattering measurement.
\cite{Murakami}
In order to investigate systems with active orbital degrees of freedom,
it is necessary to analyze the multi-orbital Hubbard model.
Such an extension of the model does not bring essential difficulty,
at least in $d$-electron materials.
For instance, in $4d$ electron systems such as ruthenate Sr$_2$RuO$_4$
introduced above, it has been found that a three $t_{\rm 2g}$-orbital
tight-binding model can well reproduce the experimental results,
as can band-structure calculations.\cite{Mazin}

However, if we try to construct such a model also for $f$-electron materials,
we immediately find a serious problem.
Since the spin-orbit interaction is fairly strong in $f$-electron systems,
the meaning of a single $f$-electron state is unclear.
Even if this point is overcome, we find another difficulty 
in the evaluation of hopping amplitudes
as well as Coulomb interactions based on a good representation for
one $f$-electron state.
In addition, we need to include correctly the important effects of
crystalline electric fields (CEF),
which are very often essential for some $f$-electron systems.

In this paper, we attempt to go beyond all those problems through use
of the $j$-$j$ coupling scheme, in order to construct an appropriate
microscopic Hamiltonian for $f$-electron compounds,
including the hopping amplitudes, Coulomb interactions, and CEF terms.
The Hamiltonian obtained is a multi-orbital Hubbard-like model.
After construction of the model, we exhibit a couple of practical
models, simplified by discarding one orbital degree of freedom.
Then, as an typical example, one simplified model is analyzed
by using exact diagonalization techniques to evaluate
several kinds of correlation functions.
Especially, the superconducting pair correlation function is carefully
analyzed based on the concept of off-diagonal long-range order.
Finally, we will discuss the relation between the present
theoretical results and some properties of CeMIn$_5$.
We also try to give a possible explanation for superconducting 
and antiferromagnetic tendencies
in ``115'' materials including CeMIn$_5$, UMGa$_5$, and PuCoGa$_5$,
based on our microscopic picture.

The organization of this paper is as follows.
In Sec.~2, the construction of the microscopic model Hamiltonian based on
the $j$-$j$ coupling scheme is discussed in detail.
The Coulomb interaction terms, CEF terms, and 
the hopping amplitudes are evaluated by paying due attention to
$f$-orbital symmetry.
Two types of simplified models are also introduced.
In Sec.~3, one of the simplified models will be analyzed 
by showing the results of spin, charge, orbital, and pair correlation
functions. A prescription to measure the pair correlation will be also
discussed.
Finally in Sec.~4, the paper is summarized and
properties of 115-materials such as CeMIn$_5$, UMGa$_5$, and
PuCoGa$_5$ are discussed.
Throughout the paper, we use units such that $\hbar$=$k_{\rm B}$=1.

%
%
\section{Model Construction}

In this section we construct a microscopic model Hamiltonian
for $f$-electron systems in order to discuss unconventional superconductivity
and novel magnetism from the microscopic viewpoint.
For this purpose, the model must include correctly the itinerant nature of
$f$ electrons as well as strong electron correlation and the effect of CEF.
Namely, the Hamiltonian $H$ should be composed of three parts as
\begin{equation}
  \label{Eq:H}
  H = H_{\rm kin}+H_{\rm int}+H_{\rm CEF},
\end{equation}
where $H_{\rm kin}$, $H_{\rm int}$, and $H_{\rm CEF}$ are $f$-electron hopping,
Coulomb interactions among $f$ electrons, and the CEF term, respectively.

In order to express the Hamiltonian in the form of Eq.~(\ref{Eq:H}), 
it is requested that the one $f$-electron state should be well-defined
by using a label $\mu$.
The physical meaning of $\mu$ will be discussed later in detail,
but here it is convenient to define a second quantized operator
to express the state $|\mu \rangle_{\bf i}$ as
\begin{equation}
 |\mu \rangle_{\bf i} = a_{{\bf i}\mu}^{\dag}|0 \rangle,
\end{equation}
where $|0 \rangle$ denotes the vacuum and
$a_{{\bf i}\mu}^{\dag}$ is the creation operator
for $f$ electrons in the $\mu$-state at site ${\bf i}$.

By using the second-quantized operator $a_{{\bf i}\mu}^{\dag}$
defined above, the kinetic term can be written as
\begin{equation}
 H_{\rm kin}=\sum_{{\bf i,a},\mu,\nu}
 t_{\mu\nu}^{\bf a} a_{{\bf i}\mu}^{\dag} a_{{\bf i+a}\nu},
\end{equation}
where $t_{\mu\nu}^{\bf a}$ is the overlap integral between
the $\mu$- and $\nu$-states connected by the vector ${\bf a}$.
The Coulomb interaction term is given in the form of
\begin{equation}
 H_{\rm int}=
 {1 \over 2}\sum_{{\bf i},\mu,\nu,\mu',\nu'}
 I(\mu,\nu;\nu',\mu')
 a_{{\bf i}\mu}^{\dag} a_{{\bf i}\nu}^{\dag}
 a_{{\bf i}\nu'} a_{{\bf i}\mu'},
\end{equation}
where $I$ is the matrix element for Coulomb interactions.
Finally, the CEF term is expressed by
\begin{equation}
 \label{H:CEF}
 H_{\rm CEF} = \sum_{{\bf i},\mu,\nu}
 B_{\mu\nu} a_{{\bf i}\mu}^{\dag} a_{{\bf i}\nu},
\end{equation}
where $B_{\mu\nu}$ indicates the potential energy for $f$ electrons
which expresses the effect of the CEF.

Now our tasks in this section are as follows:
(i) Specification of the meaning of $\mu$ for one $f$-electron state.
(ii) Evaluation of the hopping amplitude, Coulomb matrix elements,
and CEF coefficients.
(iii) Simplification of the model for practical purposes.
In the following subsections, all of these points will be clarified
in this order.

\subsection{${\bm j}$-${\bm j}$ coupling scheme}

First let us discuss the meaning of $\mu$ in the Hamiltonian Eq.~(\ref{Eq:H}).
To make it clear, it is instructive to start the discussion with
the single ion problem using two typical approaches,
the $LS$ coupling and $j$-$j$ coupling schemes.
In general, we consider the $f^n$ configuration,
where $n$ is the number of $f$ electrons included on a localized ion.
In the $LS$ coupling scheme, first the spin ${\bm S}$ and angular momentum
${\bm L}$ are formed by following Hund's rules as
${\bm S}$=$\sum_{i=1}^n {\bm s}_i$ and ${\bm L}$=$\sum_{i=1}^n {\bm \ell}_i$,
where ${\bm s}_i$ and ${\bm \ell}_i$ are spin and angular momentum
for $i$-th $f$-electron.
Note here that the Hund's rules are based on the Pauli principle and
Coulomb interactions among $f$ electrons.
After forming ${\bm S}$ and ${\bm L}$, we include the effect of spin-orbit
interaction $\lambda$${\bm L}$$\cdot$${\bm S}$, where $\lambda$ is
the spin-orbit coupling.
Note that $\lambda$$>$0 for $n$$<$7, while $\lambda$$<$0 for $n$$>$7.
Note also that a good quantum number to label such a state is the total
angular momentum ${\bm J}$, given by ${\bm J}$=${\bm L}$+${\bm S}$.
Following from simple algebra, the ground-state level is characterized by
$J$=$|L$$-$$S|$ for $n$$<$7, while $J$=$L$+$S$ for $n$$>$7.

As is easily understood from the above discussion,
the $LS$ coupling scheme is valid under the assumption that
the Hund's rule coupling is much larger than the spin-orbit interaction,
since ${\bm S}$ and ${\bm L}$ are formed by the Hund's rule coupling
prior to the inclusion of spin-orbit interaction.
It is considered that this assumption is valid for
insulating compounds with localized $f$ electrons.
However, when the spin-orbit interaction is not small compared with
the Hund's rule coupling, especially in actinide ions,
the above assumption is not always satisfied.
In addition, if the $f$ electrons become itinerant owing to hybridization
with the conduction electrons, the effect of Coulomb interactions would
thereby be effectively reduced.
In rough estimation, the effective size of the Coulomb interaction may
be as large as the bandwidth of $f$ electrons, leading to a violation of
the assumption required for the $LS$ coupling scheme.

For $f$-electron systems in which the spin-orbit interaction becomes
larger than the effective Coulomb interactions,
it is useful to consider the problem in the $j$-$j$ coupling scheme.
Here we emphasize that the $j$-$j$ coupling scheme is also quite
convenient for including many-body effects using the standard
quantum-field theoretical techniques, since individual $f$-electron states
can be clearly defined, as explained below.
First, we include the spin-orbit coupling so as to define the state
labelled by the total angular momentum ${\bm j}_i$ for the $i$-th electron,
given by ${\bm j}_i$=${\bm s}_i$+${\bm \ell}_i$.
For $f$-orbitals with $\ell$=3, we immediately obtain an octet with
$j$=7/2(=3+1/2) and a sextet with $j$=5/2(=3$-$1/2),
which are well separated by the spin-orbit interaction.
Note here that the level for the octet is higher than that of the sextet.
Then, we take into account the effect of Coulomb interactions
to accommodate $n$ electrons among the sextet and/or octet,
leading to the ground-state level in the $j$-$j$ coupling scheme.

Here, the meaning of $\mu$ in our Hamiltonian becomes clear.
In the case of $n$$<$7, $\mu$ should be the label to specify the state
in the $j$=5/2 sextet, namely,
the $z$-component of the total angular momentum
$j$=5/2 and takes the values of $\mu$=$-5/2$, $-3/2$, $\cdots$, $5/2$.
Note here that for 3$<$$n$$<$7, $j$=7/2 octet is $not$ occupied,
since we presume that the effect of spin-orbit interaction is larger
than that of the Hund's rule coupling in the $j$-$j$ coupling scheme.
On the other hand, for the case of $n$$\geq$7, $\mu$ should be
considered to specify the state in the $j$=7/2 octet,
since $j$=5/2 sextet is fully occupied.
Note again that spin-orbit interaction is larger
than that of the Hund's rule coupling.
In this paper, we concentrate only on the case of $n$$<$7,
especially the situation with $n$=1 and 2, corresponding to Ce$^{3+}$ and
U$^{4+}$ ions.
Thus, in the following, $\mu$ indicates the $z$-component of the
total angular momentum which specifies the state in the $j$=5/2 sextet.

\subsection{Coulomb interaction term}

Now let us consider the ground-state multiplet appearing in 
the $j$-$j$ coupling scheme.
For this purpose, first it is necessary to define the Coulomb interaction $I$
in $H_{\rm int}$, which is the sum of two contributions, written as
\begin{equation}
  I(\mu,\nu;\nu',\mu') = K_{\mu\nu,\nu'\mu'}-K_{\mu\nu,\mu'\nu'},
\end{equation}
where the former indicates the Coulomb term, while the latter
denotes the exchange one.
It should be noted that $I$ vanishes unless $\mu$+$\nu$=$\mu'$+$\nu'$
due to the conservation of $z$-component of total angular momentum.
The matrix element $K_{\mu_1\mu_2,\mu_3\mu_4}$ is explicitly given by
\begin{eqnarray}
  && K_{\mu_1 \mu_2,\mu_3 \mu_4} \!=\!
  \sum_{\sigma,\sigma'}  C_{\mu_1 \sigma}  C_{\mu_2 \sigma'}
  C_{\mu_3 \sigma'}  C_{\mu_4 \sigma} \nonumber \\
  && \times  \Bigl\langle \mu_1-\frac{\sigma}{2}, \mu_2-\frac{\sigma'}{2}||
  \mu_3-\frac{\sigma'}{2}, \mu_4-\frac{\sigma}{2} \Bigr\rangle,
\end{eqnarray}
where $C_{\mu\sigma}$ is the Clebsch-Gordan coefficient, given as
\begin{equation}
  C_{\mu\sigma}=-\sigma\sqrt{7/2-\mu\sigma \over 7},
\end{equation}
with $\sigma$=$+1$ ($-1$) for up (down) real spin.
The Coulomb matrix element among $\ell$-orbitals is given by
\begin{eqnarray}
  && \langle m_1, m_2 || m_3, m_4 \rangle \nonumber \\
  && \! = \! \int \! \int \! {\rm d}{\bf r}_1 \! {\rm d}{\bf r}_2
  \varphi^*_{m_1}({\bf r}_1)\varphi^*_{m_2}({\bf r}_2) g_{12}
  \varphi_{m_3}({\bf r}_2)\varphi_{m_4}({\bf r}_1),
\end{eqnarray}
where $g_{12}$=$e^2/|{\bf r}_1-{\bf r}_2|$
and $\varphi_{m}({\bf r})$ is the wavefunction for the $m$ state
in $\ell$=3 orbitals, expressed by the product of the radial function
and spherical harmonic $Y_{3 m}$.
The matrix element $\langle m_1, m_2 || m_3, m_4 \rangle$ can be written
in the form of
\begin{eqnarray}
  \langle m_1, m_2 || m_3, m_4 \rangle =
  \sum_{k=0}^{6} F^k c_k(m_1,m_4)c_k(m_2,m_3),
\end{eqnarray}
where the sum on $k$ includes only even values ($k$=0, 2, 4, and 6),
$F^k$ is the Slater-Condon parameter\cite{Slater1,Condon}
including the complex integral of the radial function,
and $c_k$ is the Gaunt coefficient,\cite{Gaunt,Racah2}
which is tabulated in the standard textbooks of quantum
mechanics (see, for instance, Ref.~\onlinecite{Slater2}).

When two electrons are accommodated in the $j$=5/2 sextet,
the allowed values for total angular momentum $J$ are 0, 2, and 4
due to the Pauli principle.
Thus, the Coulomb interaction term should be written
in a 15$\times$15 matrix form.
Note that ``15'' is the sum of the basis numbers for singlet ($J$=0),
quintet ($J$=2), and nontet ($J$=4).
As is easily understood,
this 15$\times$15 matrix can be decomposed into a block-diagonalized
form labelled by $J_z$, including one 3$\times$3 matrix for $J_z$=0,
four 2$\times$2 matrices for $J_z$=$\pm$2 and $\pm$1,
and four 1$\times$1 for $J_z$=$\pm 4$ and $\pm 3$.
We skip the details of tedious calculations for the matrix elements
and here only summarize the results in the following by using
the Racah parameters $E_k$ ($k$=0,1,2),\cite{Norman}
which are related to the Slater-Condon parameters $F^k$ as
\begin{eqnarray}
  E_0 &=& F^0-{80 \over 1225}F^2-{12 \over 441}F^4, \nonumber \\
  E_1 &=& {120 \over 1225}F^2+{18 \over 441}F^4, \\
  E_2 &=& {12 \over 1225}F^2-{1 \over 441}F^4. \nonumber
\end{eqnarray}
For $J_z$=4 and 3, we obtain
\begin{equation}
\label{Eq:Jz4}
  I(5/2,3/2;3/2,5/2)=E_0-5E_2,
\end{equation}
and
\begin{equation}
\label{Eq:Jz3}
  I(5/2,1/2;1/2,5/2)=E_0-5E_2,
\end{equation}
respectively.
For $J_z$=2 and 1, we obtain
\begin{equation}
\label{Eq:Jz2}
\begin{array}{rcl}
  I(3/2,1/2;1/2,3/2) &=& E_0+4E_2, \\
  I(5/2,-1/2;-1/2,5/2) &=& E_0, \\
  I(3/2,1/2;-1/2,5/2) &=& -3\sqrt{5}E_2,
\end{array}
\end{equation}
and
\begin{equation}
\label{Eq:Jz1}
\begin{array}{rcl}
  I(3/2,-1/2;-1/2,3/2) &=& E_0-E_2, \\
  I(5/2,-3/2;-3/2,5/2) &=& E_0+5E_2, \\
  I(3/2,-1/2;-3/2,5/2) &=& -2\sqrt{10}E_2,
\end{array}
\end{equation}
Finally, for $J_z$=0, we obtain
\begin{equation}
\label{Eq:Jz0}
\begin{array}{rcl}
  I(1/2,-1/2;-1/2,1/2) &=& E_0+2E_2+E_1, \\
  I(3/2,-3/2;-3/2,3/2) &=& E_0-3E_2+E_1, \\
  I(5/2,-5/2;-5/2,5/2) &=& E_0+5E_2+E_1, \\
  I(1/2,-1/2;-3/2,3/2) &=& -E_1-3E_2,  \\
  I(1/2,-1/2;-5/2,5/2) &=& E_1-5E_2, \\
  I(3/2,-3/2;-5/2,5/2) &=& -E_1.
\end{array}
\end{equation}
Note here the following relations:
\begin{equation}
 I(\mu,\nu;\nu',\mu')=I(\mu',\nu';\nu,\mu),
\end{equation}
and
\begin{equation}
 I(\mu,\nu;\nu',\mu')=I(-\nu,-\mu;-\mu',-\nu').
\end{equation}
By using these two relations and Eqs.~(\ref{Eq:Jz4}-\ref{Eq:Jz0}), 
we can obtain all the Coulomb matrix elements.\cite{Inglis}

As a typical example, let us consider the $f^2$ configuration.
In the $j$-$j$ coupling schemes,
two electrons are accommodated in the $j$=5/2 sextet.
When we diagonalize the 15$\times$15 matrix for Coulomb interaction terms,
we can easily obtain the eigen energies as
$E_0$$-$$5E_2$ for the $J$=4 nontet,
$E_0$+$9E_2$ for the $J$=2 quintet,
and $E_0$+$3E_1$ for the $J$=0 singlet.\cite{Norman}
Since the Racah parameters are all positive, the ground state is specified
by $J$=4 in the $j$-$j$ coupling scheme.
In the $LS$ coupling scheme, on the other hand, we obtain the ground-state
level as $^{3}H$ with $S$=1 and $L$=5 from the Hund's rules.
On further inclusion of the spin-orbit interaction,
the ground state becomes characterized by $J$=4,
expressed as $^3H_4$ in the traditional notation.
Note that we are now considering a two-electron problem.
Thus, if we correctly include the effects of Coulomb interactions,
it is concluded that the same quantum number as that in
the $LS$ coupling scheme is obtained in the $j$-$j$ coupling scheme
for the ground-state multiplet.

In order to understand the physical meaning of Racah parameters,
let us consider a simplified Coulomb interaction term.
In the above discussion, the expressions using Racah parameters
are not convenient since they depend on the orbitals
in a very complicated manner, although they keep the correct symmetry
required by group theory.
To clarify their meanings, it may be useful to restart the discussion from
the following interaction form among $\ell=3$ orbitals:
\begin{eqnarray}
  \label{Eq:Hint2}
  H_{\rm int} &=&
  U \sum_{{\bf i}m}\rho_{{\bf i}m\uparrow} \rho_{{\bf i}m\downarrow}
  + U'/2 \sum_{{\bf i},\sigma,\sigma',m \ne m'} 
  \rho_{{\bf i}m\sigma} \rho_{{\bf i}m'\sigma'}
  \nonumber \\
  &+& J/2 \sum_{{\bf i},\sigma,\sigma',m \ne m'} 
  f_{{\bf i}m\sigma}^{\dag} f_{{\bf i}m'\sigma'}^{\dag}
  f_{{\bf i}m\sigma'}f_{{\bf i}m'\sigma},
\end{eqnarray}
where $f_{{\bf i}m\sigma}$ is the annihilation operator
for an $f$ electron with spin $\sigma$ in the $m$ state
of angular momentum $\ell$ at site {\bf i} and
$\rho_{{\bf i}m\sigma}$=
$f_{{\bf i}m\sigma}^{\dag} f_{{\bf i}m\sigma}$.
In this equation, we include only three interactions;
intra-orbital Coulomb interaction $U$, inter-orbital Coulomb interaction $U'$,
and the exchange interaction $J$.
Note that the relation $U$=$U'$+$J$ 
holds among Coulomb interactions to ensure
rotational invariance in the orbital space.

By using Clebsch-Gordan coefficients, $f_{{\bf i}m\sigma}$
with real-spin $\sigma$ can be related to $a_{{\bf i}\mu}$ as
\begin{equation}
  f_{{\bf i}m\sigma}=-\sigma\sqrt{3-\sigma m \over 7}a_{{\bf i}m+\sigma/2}.
\end{equation}
Note here that we consider only the $j=5/2$ sextet.
The Coulomb interaction term for $j$=5/2 is given by
\begin{eqnarray}
  H_{\rm int} =
  U_{\rm eff} \sum_{{\bf i} \mu>\mu'}
  n_{{\bf i}\mu} n_{{\bf i}\mu'}
  -J_{\rm H} {\bm J}_{\bf i}^2
  +(35J_{\rm H}/4)N_{\bf i},
\end{eqnarray}
where $n_{{\bf i}\mu}$=$a_{{\bf i}\mu}^{\dag}a_{{\bf i}\mu}$,
$N_{\bf i}$=$\sum_{\mu}n_{{\bf i}\mu}$,
$U_{\rm eff}$=$U'$$-$$J/2$,
$J_{\rm H}$=$J/49$,
and ${\bm J}_{\bf i}$ is the operator for total angular momentum
with $J$=5/2.
Explicitly, ${\bm J}_{\bf i}^2$ is written as
\begin{eqnarray}
 {\bm J}_{\bf i}^2 &=& \sum_{\mu,\mu'}
 [\mu\mu' n_{{\bf i}\mu}n_{{\bf i}\mu'}
   +(\phi^{+}_{\mu}\phi^{-}_{\mu'}
 a_{{\bf i}\mu+1}^{\dag} a_{{\bf i}\mu}
 a_{{\bf i}\mu'-1}^{\dag}a_{{\bf i}\mu'} \nonumber \\
 &+& \phi^{-}_{\mu}\phi^{+}_{\mu'}
 a_{{\bf i}\mu-1}^{\dag} a_{{\bf i}\mu}
 a_{{\bf i}\mu'+1}^{\dag}a_{{\bf i}\mu'})/2],
\end{eqnarray}
with $\phi_{\mu}^{\pm}=\sqrt{j(j+1)-\mu(\mu\pm 1)}$
=$\sqrt{35/4-\mu(\mu\pm 1)}$.

For two electrons in the $j$=5/2 sextet, based upon
the simplified Coulomb interaction term, we can easily obtain
the energy levels as $U_{\rm eff}$$-$$5J_{\rm H}/2$ for the $J$=4 nontet,
$U_{\rm eff}$+$23J_{\rm H}/2$ for the $J$=2 quintet,
and $U_{\rm eff}$+$35J_{\rm H}/2$ for the $J$=0 singlet.
When we compare these energy levels with the results
obtained using Racah parameters,
we understand the correspondence such as
$E_0$$\sim$$U_{\rm eff}$ and $E_2$$\sim$$J_{\rm H}$.
Namely, $E_0$ is the effective inter-orbital Coulomb interaction,
while $E_2$ denotes the Hund's rule coupling.
Note that $E_1$ does not appear, since it is related to the
pair-hopping interaction which is not included here.

Finally, we note the smallness of $J_{\rm H}$, given as $J_{\rm H}$=$J/49$.
The origin of the large reduction factor 1/49 is,
in one word, due to the neglect of $j$=7/2 octet.
In the Coulomb interaction term Eq.~(\ref{Eq:Hint2}),
the Hund's rule term is simply written as $-J {\bm S}^2$. 
Note the relation ${\bm S}$=$(g_J-1){\bm J}$ with $g_J$
the Land\'e's $g$-factor.
For $j$=5/2, we easily obtain $g_J$=6/7, indicating ${\bm S}$=$-(1/7){\bm J}$.
Thus, the original Hund's rule term is simply rewritten as $-(J/49) {\bm J}^2$.

\subsection{CEF term}

As is well known, the ground-state multiplet is further split due to the effect
of the CEF.
In this paper we take into account the effect of the CEF by simply accommodating
several electrons in the one-electron potential.
Namely, $B_{\mu\nu}$ in Eq.~(\ref{H:CEF}) is expressed by the CEF parameters
for $J$=5/2.
Since the CEF term depends on the crystal structure,
it is convenient to write down the CEF formula for $J$=5/2 
in the typical lattice structure, expressed as
\begin{equation}
  H_{\rm CEF}^{\rm cub}=
   B_4^0({\hat O}_4^0+5{\hat O}_4^4),
\end{equation}
for a cubic lattice,
\begin{equation}
  H_{\rm CEF}^{\rm tet}=
   B_2^0 {\hat O}_2^0+B_4^0 {\hat O}_4^0
  +B_4^4 {\hat O}_4^4,
\end{equation}
for a tetragonal lattice, and
\begin{equation}
  H_{\rm CEF}^{\rm hex}=
   B_2^0 {\hat O}_2^0+B_4^0 {\hat O}_4^0,
\end{equation}
for a hexagonal lattice.
In each equation above, ${\hat O}_n^m$ is the Stevens equivalent
operator,\cite{Stevens} composed of $J_z$, $J_+$, and $J_-$.
The matrix formulae have been tabulated in a paper by
Hutchings.\cite{Hutchings}
The coefficients $B_n^m$ are conventionally called the CEF parameters,
which are, in actuality, determined by the fitting of experimental
results for physical quantities such as magnetic susceptibility.

Note here that the above formulae have been obtained from the case of $J$=5/2.
In general, the CEF term is expressed in matrix form, depending on the value of $J$;
for $J$ larger than 5/2, higher terms in $B_n^m$ should occur.
However, as already mentioned above, since in this paper the effect of the CEF is
considered as a one-electron potential based on the $j$-$j$ coupling scheme,
it is enough to use the CEF term for $J$=5/2.

In order to check the validity of our viewpoint regarding
the CEF term, let us see, for instance, the case of cubic structure,
in which we can easily obtain
\begin{equation}
\label{Eq:cubic-CEF}
\begin{array}{rcl}
  B_{\pm 5/2,\pm 5/2} &=&  60 B_4^0, \\
  B_{\pm 3/2,\pm 3/2} &=& -180 B_4^0, \\
  B_{\pm 1/2,\pm 1/2} &=&  120 B_4^0, \\
  B_{\pm 5/2,\mp 3/2} &=& B_{\mp 3/2,\pm 5/2}=60\sqrt{5} B_4^0,
\end{array}
\end{equation}
and zero for other $\mu$ and $\nu$.
In the following we consider the cases of both $f^1$ and $f^2$ configurations.

When one electron exists in $f$-orbitals, i.e., in the case of
the Ce$^{3+}$ ion, the ground state is
easily obtained by the diagonalization of Eq.~(\ref{Eq:cubic-CEF}).
Then we immediately obtain two eigen energies,
$E(\Gamma_7)$=$-240B_4^0$ for the $\Gamma_7$ doublet and
$E(\Gamma_8)$=$120B_4^0$ for the $\Gamma_8$ quartet.
Thus, the ground state is changed due to the sign of $B_4^0$,
as shown in Fig.~1(a).
Qualitatively, the ground state is deduced from the positions of
ligand anions and the shapes of wavefunctions.
As in the case of the Ce monopnictide,
when anions are located along the direction of [1,0,0] from
the Ce$^{3+}$ ion, the $\Gamma_8$ level is energetically penalized,
since the corresponding wavefunctions are expanding along the
axis directions.
On the other hand, as in the case of Ce hexaborides
with anions located along the [1,1,1] directions from the Ce$^{3+}$ ions,
the $\Gamma_7$ levels are higher due to the shape of 
$\Gamma_7$ wavefunction.

\begin{figure}
\includegraphics[width=1.0\linewidth,height=7.0truecm]{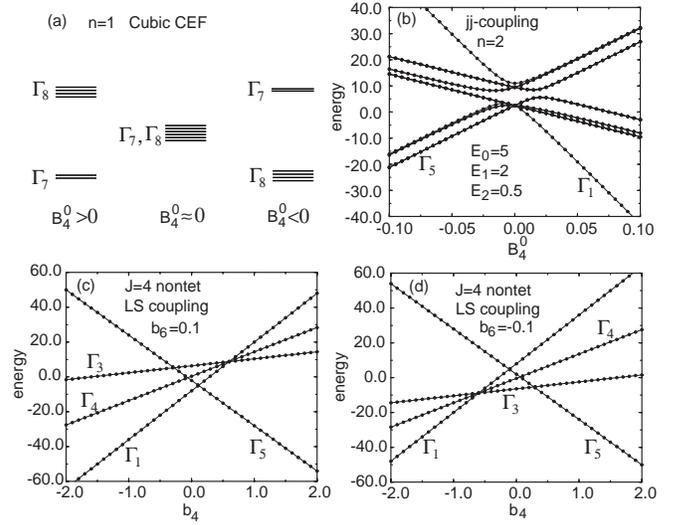}
\caption{
(a) Schematic view of level scheme for $n$=1 under the cubic CEF.
(b) Energy levels as a function of $B_4^0$ for $n$=2 in the $j$-$j$
coupling scheme.
(c) Energy levels as a function of $b_4$(=$60B_4^0$) for $J$=4
nontet under the cubic CEF with $b_6$(=$1260B_6^0$)=0.1.
(d) Energy levels as a function of $b_4$ for $J$=4 nontet
with $b_6$=$-0.1$.}
\end{figure}

In order to consider the $f^2$ configurations such as Pr$^{4+}$
and U$^{4+}$ ions based on the $j$-$j$ coupling scheme,
we put two electrons in the potential for the $J$=5/2 CEF term.
Here some readers may have a naive question regarding its validity.
To clarify this point, we analyze the CEF level schemes for the
$f^2$ configuration by diagonalizing $H_{\rm int}$+$H_{\rm CEF}$.
In Fig.~1(b), we show the eigen energies as a function of $B_4^0$
for a set of typical values of Racah parameters,
$E_0$=5, $E_1$=2, and $E_2$=0.5.
For $B_4^0$$<$0, the ground state is the $\Gamma_5$ triplet, while
the $\Gamma_1$ singlet appears for $B_4^0$$>$0. 
At $B_4^0$=0, the ground-state is the $J$=4 nontet.
Here it should be noted that the ground-state wavefunction
consists of a mixture of the $J$=4, $J$=2, and $J$=0 states.
Note also that $J_z$, not $J$, is a good quantum number to
label the state, since the rotational invariance in $J$ space
is broken by the effect of the CEF for $J$=5/2.
However, we confirm that the contribution of the $J$=4 component is
the largest in the ground state wave function.

In the $LS$ coupling scheme, on the other hand,
we need to consider a 9$\times$9 matrix form for the $J$=4 nontet.
The CEF term for $J$=4 under cubic symmetry is given by
$H_{\rm CEF}^{\rm cub}$=
$B_4^0({\hat O}_4^0+5{\hat O}_4^4)$+$
B_6^0({\hat O}_6^0-21{\hat O}_6^4)$.
Note that for the case of $J$=4, there appears a higher order term
including $B_6^0$.
Referring to the paper by Hutchings,\cite{Hutchings}
we obtain the CEF coefficients as
\begin{equation}
\begin{array}{rcl}
  B_{\pm 2,\pm 2} &=& -660 (B_4^0-42 B_6^0), \\
  B_{\pm 2,\mp 2} &=&  180 (5B_4^0+ 294 B_6^0),
\end{array}
\end{equation}
for the $J_z$=$\pm$2 sector,
\begin{equation}
\begin{array}{rcl}
  B_{\pm 3,\pm 3} &=& -1260 (B_4^0+17 B_6^0), \\
  B_{\mp 1,\mp 1} &=&  180 (3 B_4^0+7 B_6^0), \\
  B_{\pm 3,\mp 1} &=& B_{\pm 1,\mp 3}=
  60 \sqrt{7} (5 B_4^0 + 63 B_6^0),
\end{array}
\end{equation}
for the $J_z$=$\pm$3 and $\mp$1 sectors, and
\begin{equation}
  \begin{array}{rcl}
  B_{\pm 4,\pm 4} &=&  840 (B_4^0+ 6 B_6^0), \\
  B_{0,0} &=&  360 (3 B_4^0- 7 B_6^0), \\
  B_{\pm 4,0} &=& B_{0,\pm 4} =
  60 \sqrt{70} (B_4^0 - 126 B_6^0),
\end{array}
\end{equation}
for the $J_z$=$\pm$4 and 0 sectors.
Note that we obtain three 2$\times$2 and one 3$\times$3 matrices.
After diagonalization, we obtain four eigen energies as
\begin{equation}
\begin{array}{rcl}
  E(\Gamma_1) &=& 1680 (B_4^0- 60 B_6^0) ~~ ({\rm singlet}), \\
  E(\Gamma_3) &=& 240 (B_4^0+ 336 B_6^0) ~~ ({\rm doublet}), \\
  E(\Gamma_4) &=& 840 (B_4^0+ 6 B_6^0) ~~ ({\rm triplet}), \\
  E(\Gamma_5) &=& -120 (13 B_4^0+210 B_6^0) ~~ ({\rm triplet}).
\end{array}
\end{equation}
As shown in Fig.~1(c),
for $B_6^0$$\agt$0, the ground state is a $\Gamma_5$ triplet for $B_4^0$$>$0,
while it is a $\Gamma_1$ singlet for $B_4^0$$<$0.
On the other hand, for $B_6^0$$<$0, the ground state is also
a $\Gamma_5$ triplet for $B_4^0$$>$0 and a $\Gamma_1$ singlet for $B_4^0$$<$0,
respectively, except for the region around $B_4^0$$\sim$0, in which case
the doublet $\Gamma_3$ becomes the ground state, as shown in Fig.~1(d).

When we compare the $LS$ coupling results with those for the $j$-$j$
coupling, the following two points are noted:
(i) Although we commonly observe conversion of the ground state
between a $\Gamma_5$ triplet and a $\Gamma_1$ singlet,
the correspondence with the sign of $B_4^0$ is reversed
between the $LS$ and the $j$-$j$ coupling schemes.
(ii) The $\Gamma_3$ doublet does $not$ appear as the ground state
in the $j$-$j$ coupling scheme.
The point (i) may not be serious, since it is just a problem
of parametrization.
However, point (ii) looks strange at first glance.
In order to clarify this point, it is useful to change the
basis in the $j$-$j$ coupling.
As shown in Fig.~1(a), in the one $f$-electron case, for $B_4^0$$>$0 ($<$0),
the $\Gamma_7$ doublet is the ground state (excited state),
while the $\Gamma_8$ quartet is the excited state (ground state).
If we introduce pseudo-spin to distinguish the Kramers doublet,
$\Gamma_7$ indicates a single orbital,
while $\Gamma_8$ denotes doubly degenerate orbitals.
When we put two electrons into the situation shown in Fig.~1(a),
owing to the Hund's rule coupling we obtain three cases:
(i) Two electrons occupy $\Gamma_7$ for $B_4^0$$>$0.
(ii) They occupy $\Gamma_8$ for $B_4^0$$<$0.
(iii) One occupies $\Gamma_7$ and the other does $\Gamma_8$
around $B_4^0$$\approx$0.
After algebraic calculations, we confirm that (i) and (ii)
correspond to a $\Gamma_1$ singlet and a $\Gamma_5$ triplet,
respectively, in agreement with the above result.
However, in contrast to the naive expectation, (iii) results in
a $\Gamma_4$ triplet due to the Hund's rule coupling.
Note that this $\Gamma_4$ triplet does $not$ become the
unique ground state, as shown in Fig.~1(b), except at $B_4^0$=0.
In this case, the $\Gamma_3$ doublet appears only as an excited state,
since this $\Gamma_3$ doublet is found to be composed of two
singlet states, each of which includes a couple of electrons,
one in $\Gamma_7$ and the other in $\Gamma_8$.
As is easily understood, a local singlet state in multi-orbital
systems is energetically penalized due to the Hund's rule coupling,
and thus, the $\Gamma_3$ doublet is always the excited state,
higher than the $\Gamma_4$ triplet.

As long as we consider the $j$-$j$ coupling scheme for the $j$=5/2 sextet,
the $\Gamma_3$ doublet never appears as the ground state,
which is attributed to ignoring the $j$=7/2 octet.
However, it is $not$ allowed to include the effect of this octet
in the present $j$-$j$ coupling scheme, since it just contradicts
the original assumption.
Namely, if the Hund's rule coupling is strong enough so that $j$=7/2 cannot
be neglected, we must $not$ consider the $j$-$j$ coupling schemes,
but take the $LS$ coupling scheme from the outset.
Here we have an interesting conclusion:
In the $j$-$j$ coupling scheme, $\Gamma_3$ never becomes
the ground state. Only in the $LS$ coupling scheme with
$B_6^0$$<$0 and $B_4^0$$\approx$0, the ground state becomes $\Gamma_3$.
Except for the disappearance of the $\Gamma_3$ doublet, it is concluded that
the CEF terms can be treated within the $j$-$j$ coupling schemes.

\subsection{Hopping term}

Now we consider the hopping motion of $f$ electrons based on the $j$-$j$ coupling scheme.
As discussed in the previous subsections, the one $f$-electron state
is labelled by $\mu$.
Within the second-order perturbation,
the hopping amplitude $t^{\bf a}_{\mu\nu}$
includes two contributions, as shown in Fig.~2:
(a) The direct hopping of $f$ electrons between the $\mu$-state
at ${\bf i}$ site and the $\nu$-state at ${\bf i+a}$ site
and (b) the hopping via ligand anions located
along the ${\bf a}$-direction from the $f$-ion site.
In order to evaluate both contributions,
basically we employ Slater's two-center integral method,
\cite{Slater-Koster}
as has been done in $d$-electron systems.
In fact, the table for two-center integrals has been extended
to the case including $f$ orbitals by Takegahara {\it et al.}
\cite{Takegahara}
However, as emphasized in the previous subsections, it is necessary to
pay due attention to the fact that spin and orbital are mixed.
Thus, in this case, it is $not$ allowed to simply use the
Slater-Koster table.

\begin{figure}
\includegraphics[width=1.0\linewidth,height=4.2truecm]{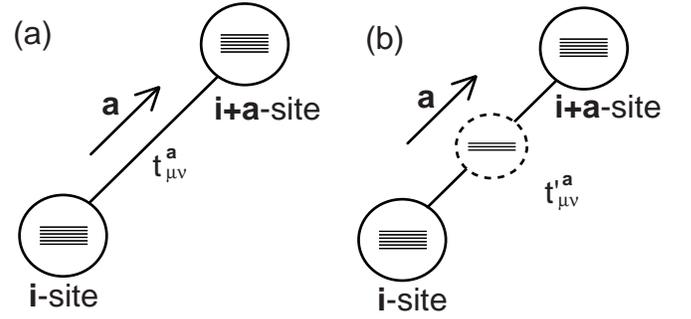}
\caption{
Schematic views for (a) direct $f$-electron hopping $t_{\mu\nu}^{\bf a}$
between neighboring sites along ${\bf a}$-direction and (b)
hopping amplitude ${t'}_{\mu\nu}^{\bf a}$ via ligand anions.
}
\end{figure}

To avoid any confusion, it is convenient to step back to $f$-electron operators
in the $\ell$=3 multiplet, defined as $f_{{\bf i}m\sigma}$.
Using Clebsch-Gordan coefficients, $f_{{\bf i}m\sigma}$
with real-spin $\sigma$ is related to $a_{{\bf i}\mu}$ as
\begin{equation}
  a_{{\bf i}\mu}=\sum_{\sigma} C_{\mu\sigma}
  f_{{\bf i},\mu-\sigma/2,\sigma}.
\end{equation}
Note again here that 
$\sigma$ is defined to be $+1$ ($-1$) for up (down) spin.

First, let us consider the direct hopping of $f$ electrons
along the ${\bf a}$-direction, as shown in Fig.~2(a).
Since the real spin should be conserved in the hopping process,
$t^{\bf a}_{\mu\nu}$ is given as
\begin{equation}
 \label{Eq:hopping1}
 t^{\bf a}_{\mu\nu} = \sum_{\sigma}
 C_{\mu\sigma}C_{\nu\sigma} T^{\bf a}_{3,\mu-\sigma/2;3,\nu-\sigma/2},
\end{equation}
where $T^{\bf a}_{\ell,m; \ell',m'}$ is the hopping amplitude of
electrons between $(\ell,m)$- and $(\ell',m')$-states along
the ${\bf a}$-direction.

Now the problem is reduced to the evaluation of $T^{\bf a}_{\ell,m; \ell',m'}$.
Although we can simply consult the paper of Slater and Koster,\cite{Slater-Koster}
a convenient formula has been obtained by Sharma for the overlap integral between
two orbitals, $(\ell,m)$ and $(\ell',m')$,\cite{Sharma} connected by unit vector
${\bf a}$. It is expressed as
\begin{equation}
 \label{Eq:hopping2}
 T^{\bf a}_{\ell,m;\ell',m'}
 \!=\! (\ell \ell' \sigma) \sqrt{4\pi \over 2\ell+1} \sqrt{4\pi \over 2\ell'+1}
 Y^{*}_{\ell m}(\theta,\varphi) Y_{\ell' m'}(\theta,\varphi),
\end{equation}
where $(\ell \ell' \sigma)$ denotes Slater's two-center integral
through the $\sigma$ bond, for instance, it is
$(ff\sigma)$ for $\ell$=$\ell'$=3
and $(fp\sigma)$ for $\ell$=3 and $\ell'$=1.
$Y_{\ell m}$ is the spherical harmonic function with
$m$=$-\ell$, $\cdots$, $\ell$.
$\theta$ and $\varphi$ are polar and azimuth angles, respectively,
to specify the vector {\bf a} as
\begin{equation}
{\bf a}=
(\sin \theta \cos \varphi, \sin \theta \sin \varphi, \cos \theta).
\end{equation}

Let us consider the direct hopping between $f$ orbitals
in nearest neighbor sites by putting $\ell$=$\ell'$=3.
After some calculations, we obtain the hopping amplitudes
as follows. For diagonal elements, we obtain
\begin{equation}
\begin{array}{rcl}
 t^{\bf a}_{\pm 5/2,\pm 5/2} &=& 5t_0 \sin^4 \theta, \\
 t^{\bf a}_{\pm 3/2,\pm 3/2} &=& t_0 \sin^2 \theta (1+15 \cos ^2 \theta), \\
 t^{\bf a}_{\pm 1/2,\pm 1/2} &=& 2t_0 (1-2 \cos ^2 \theta +5 \cos^4 \theta),
\end{array}
\end{equation}
where the energy unit $t_0$ is given by
\begin{equation}
 t_0={3 \over 56}(ff\sigma),
\end{equation}
where $(ff\sigma)$ is the Slater-Koster two-center integral
between adjacent $f$ orbitals. Note that $t^{\bf a}_{\mu,-\mu}=0$.
For off-diagonal elements, we obtain
\begin{equation}
\begin{array}{rcl}
 t^{\bf a}_{\pm 5/2, \pm 1/2} &=&
 -t_0 \sqrt{10}e^{\mp 2i\varphi} \sin^2 \theta (1 - 3 \cos^2 \theta), \\
 t^{\bf a}_{\pm 5/2, \mp 3/2} &=& t_0 \sqrt{5}e^{\mp 4i\varphi} \sin^4 \theta, \\
 t^{\bf a}_{\pm 1/2, \mp 3/2} &=& -t_0 \sqrt{2}e^{\mp 2i\varphi} \sin^2 \theta
 (1 + 5 \cos^2 \theta),
\end{array}
\end{equation}
and
\begin{equation}
\begin{array}{rcl}
 t^{\bf a}_{5/2,-1/2} &=& -t^{\bf a}_{1/2,-5/2}=
 t_0 \sqrt{10}e^{-3i\varphi} \sin^2 \theta \sin 2\theta, \\
 t^{\bf a}_{5/2, 3/2} &=& -t^{\bf a}_{-3/2, -5/2}=
 -2t_0 \sqrt{5}e^{-i\varphi} \sin^2 \theta \sin 2\theta, \\
 t^{\bf a}_{1/2, 3/2} &=& -t^{\bf a}_{-3/2, -1/2}=
 t_0 \sqrt{2}e^{i\varphi} \sin 2\theta 
 (1 - 5 \cos^2 \theta). 
\end{array}
\end{equation}
Note that $t^{\bf a}_{\nu\mu}$=$t^{{\bf a}*}_{\mu\nu}$.

Next we consider the hopping of $f$ electrons ${t'}^{\bf a}_{\mu\nu}$
through ligand anions, as shown in Fig.~2(b).
Assume that $\ell$-orbitals are situated along the
${\bf a}$-direction and $f$-electrons hop to the neighboring $f$-site
through these $\ell$-orbitals.
Again we note that the real spin is conserved in the hopping process.
In second-order perturbation, we can easily obtain
\begin{eqnarray}
 {t'}^{\bf a}_{\mu\nu} =
 \sum_{m,\sigma} C_{\mu\sigma}C_{\nu\sigma}
 {T^{\bf a}_{3,\mu-\sigma/2;\ell,m} T^{\bf a}_{\ell,m;3,\nu-\sigma/2} 
 \over \varepsilon_{\rm f}-\varepsilon_{\ell}},
\end{eqnarray}
where $\varepsilon_{\ell}$ and $\varepsilon_{\rm f}$ are
the energy levels for $\ell$- and $f$-orbitals and
$m$ runs between $-\ell$ and $\ell$.
Due to the orthogonality of spherical harmonics, it is easily reduced to
\begin{equation}
 {t'}^{\bf a}_{\mu\nu}={(f \ell \sigma)^2 \over
 \varepsilon_{\rm f}-\varepsilon_{\ell}}
 {t^{\bf a}_{\mu\nu} \over (ff\sigma)}.
\end{equation}
It is stressed that
the ${\bf a}$-dependence of ${t'}^{\bf a}_{\mu\nu}$ is the same as that
of ${t}^{\bf a}_{\mu\nu}$,
although the energy unit is changed from $(ff\sigma)$
to $(f \ell \sigma)^2/(\varepsilon_{\rm f}-\varepsilon_{\ell})$.
Note that in general, the sign is also changed.

\subsection{Simplified models}

In previous subsections, we have completed the construction of a model Hamiltonian
which is expected to be a basic model to investigate the microscopic
properties of $f$-electron systems.
However, it includes six states per site, i.e., three Kramers doublets.
For practical purposes, it is more convenient to simplify
the model further by discarding one of the three Kramers doublets.
Although the hopping amplitudes are obtained in a general form
applicable to any type of crystal structure,
it is necessary to set the crystal structure
for the purpose of actual calculations.

\subsubsection{$\Gamma_8$ model}

First let us consider a simple cubic lattice composed of ions
with $f$ electrons.
Under the cubic CEF, as discussed above,
for $B_4^0$$<$0 the $\Gamma_8$ quartet becomes the ground state,
while the $\Gamma_7$ doublet is the excited state.
This situation is considered to be realized when ligand anions
are situated in the face-centered or body-centered position,
due to a naive discussion of the point-charge picture.
Note that the effect of ligand anions is included only through
the CEF term, and we do not consider its effect on $f$-electron hopping.
Then, it is possible to consider a model consisting of $\Gamma_8$,
neglecting $\Gamma_7$ under the condition of $B_4^0$$<$0.

To consider the model including $\Gamma_8$ levels,
it is useful to define new operators with ``orbital'' degrees of
freedom to distinguish two Kramers doublets included in $\Gamma_8$ as
\begin{equation}
\begin{array}{rcl}
 c_{{\bf i}\alpha \uparrow}
 &=& \sqrt{5/6}a_{{\bf i}-5/2}+\sqrt{1/6}a_{{\bf i}3/2}, \\
 c_{{\bf i}\alpha \downarrow}
 &=& \sqrt{5/6}a_{{\bf i}5/2}+\sqrt{1/6}a_{{\bf i}-3/2},
\end{array}
\end{equation}
for $\alpha$ orbital electrons and 
\begin{equation}
 c_{{\bf i}\beta \uparrow}=a_{{\bf i}-1/2},~~
 c_{{\bf i}\beta \downarrow}=a_{{\bf i}1/2},
\end{equation}
for $\beta$ orbital electrons, respectively.
Note that the $\Gamma_7$ state is given as a $\gamma$ orbital as
\begin{equation}
\begin{array}{rcl}
 c_{{\bf i}\gamma \uparrow}
 &=& \sqrt{1/6}a_{{\bf i}-5/2}-\sqrt{5/6}a_{{\bf i}3/2}, \\
 c_{{\bf i}\gamma \downarrow}
 &=& \sqrt{1/6}a_{{\bf i}5/2}-\sqrt{5/6}a_{{\bf i}-3/2}.
\end{array}
\end{equation}
For the standard time reversal operator ${\cal K}$=$-{\rm i}\sigma_y K$,
where $K$ denotes an operator to take the complex conjugate,
we can easily show the relation
\begin{equation}
 {\cal K}c_{{\bf i}\tau \sigma}=\sigma c_{{\bf i}\tau -\sigma}.
\end{equation}
Note that this has the same definition for real spin.

For hopping in the $xy$ plane and along the $z$-axis,
we simply set ($\theta$, $\varphi$) to be ($\pi/2$,0), ($\pi/2$,$\pi/2$),
and (0,0) for ${\bf a}$=${\bf x}$=[1,0,0],
${\bf y}$=[0,1,0], and 
${\bf z}$=[0,0,1], respectively.
Then, by using the general results in the previous subsection,
we easily obtain 
$t^{\bf a}_{\mu\nu}$ between neighboring $f$ orbitals
in the $xy$ plane and along the $z$ axis.
Further we transform the basis by the above definitions
for operators with orbital degrees of freedom.
The results are given as\cite{Maehira1,Takimoto}
\begin{equation}
  t_{\tau\tau'}^{\bf x} = t
\left(
\begin{array}{ccc}
1 & -1/\sqrt{3} & 0 \\
-1/\sqrt{3} & 1/3 & 0 \\
0 & 0 & 0 \\
\end{array}
\right),
\end{equation}
for the ${\bf x}$-direction,
\begin{equation}
  t_{\tau\tau'}^{\bf y} = t
\left(
\begin{array}{ccc}
1 & 1/\sqrt{3} & 0 \\
1/\sqrt{3} & 1/3 & 0 \\
0 & 0 & 0 \\
\end{array}
\right),
\end{equation}
for the ${\bf y}$ direction, and
\begin{equation}
  t_{\tau\tau'}^{\bf z} = t
\left(
\begin{array}{ccc}
0 & 0 & 0 \\
0 & 4/3 & 0 \\
0 & 0 & 0 \\
\end{array}
\right),
\end{equation}
for the ${\bf z}$ direction. Note that $t$=$6t_0$=(9/28)$(ff\sigma)$.
Here we note two points:
(i) Hopping amplitudes among $\Gamma_8$ orbitals are just the
same as those for the $e_{\rm g}$ orbitals of $3d$ electrons.\cite{Hotta2}
This is quite natural if we recall the fact that
$\Gamma_8$ is isomorphic to $\Gamma_3 \times \Gamma_6$,
where $\Gamma_3$ indicates $E$ representation for the orbital part
and $\Gamma_6$ denotes the spin part.
(ii) The $\Gamma_7$ orbital is localized, since the corresponding wavefunction
has nodes along the axis direction.
Since $\Gamma_7$ is the excited state for $B_4^0$$<$0 and it is also localized,
electrons do not occupy the $\Gamma_7$ orbital.

After lengthy and tedious calculations to transform the basis of
the Coulomb interaction terms, the Hamiltonian for $\Gamma_8$ is
given by\cite{Hotta1}
\begin{eqnarray}
H_8 &=&
\sum_{{\bf i,a},\sigma,\tau,\tau'} t^{\bf a}_{\tau\tau'}
c^{\dag}_{{\bf i}\tau\sigma}f_{{\bf i+a}\tau'\sigma}
-\varepsilon \sum_{\bf i}(\rho_{{\bf i}a}-\rho_{{\bf i}b})/2 \nonumber \\
&+&U \sum_{{\bf i},\tau}\rho_{{\bf i}\tau\uparrow} \rho_{{\bf i}\tau\downarrow}
+U' \sum_{\bf i} \rho_{{\bf i}a} \rho_{{\bf i}b}
\nonumber \\
&+& J/2 \sum_{{\bf i},\sigma,\sigma'}
\sum_{\tau \ne \tau'} 
c_{{\bf i}\tau\sigma}^{\dag} c_{{\bf i}\tau'\sigma'}^{\dag}
c_{{\bf i}\tau\sigma'} c_{{\bf i}\tau'\sigma} \nonumber \\
&+& J' \sum_{{\bf i},\tau \ne \tau'}
c_{{\bf i}\tau\uparrow}^{\dag} c_{{\bf i}\tau'\downarrow}^{\dag}
c_{{\bf i}\tau\downarrow} c_{{\bf i}\tau'\uparrow},
\end{eqnarray}
where
$\rho_{{\bf i}\tau\sigma}$=
$c_{{\bf i}\tau\sigma}^{\dag} c_{{\bf i}\tau\sigma}$
and
$\rho_{{\bf i}\tau}$=$\sum_{\sigma}\rho_{{\bf i}\sigma\tau}$.
Note that in the second term, the level splitting $\varepsilon$ is
simply introduced to include partially the effect of a tetragonal CEF.
In the Coulomb interaction terms, $U$, $U'$, $J$, and $J'$ denote
intra-orbital, inter-orbital, exchange, and pair-hopping interactions, respectively,
expressed by the Racah parameters $E_k$ as
\begin{equation}
\begin{array}{rcl}
 U &=& E_0+E_1+2E_2, \\
 U' &=& E_0+(2/3)E_2, \\
 J &=& 5E_2, \\
 J'&=& E_1-(11/3)E_2.
\end{array}
\end{equation}
Note that the relation $U$=$U'$+$J$+$J'$  holds, ensuring rotational
invariance in pseudo-orbital space for the interaction part.

We believe that this $\Gamma_8$ Hamiltonian provides a simple, but non-trivial
model to consider superconductivity and magnetism in $f$-electron systems.
Note again that it is essentially the same as the model for
$e_{\rm g}$ electron systems such as manganites,\cite{Hotta2}
although the coupling with Jahn-Teller distortion is not included
in the present model.
Due to the complex interplay and competition among charge, spin, and orbital
degrees of freedom, a rich phase diagram has been obtained for manganites.
\cite{Dagotto2}
Thus, it is definitely expected that a similar richness will also be
unveiled for $f$-electron systems based on the $\Gamma_8$ model Hamiltonian.

\subsubsection{$\Gamma_7$ model}

The $\Gamma_8$ Hamiltonian $H_8$ is believed to be a non-trivial model for
$f$-electron systems, since we can immediately grasp some essential aspects of
$f$-electron materials viewed as charge-spin-orbital complex systems.
In fact, based on $H_8$, we can have insight into the effect on superconductivity
of cooperation and competition among spin and orbital fluctuations 
in the new family of heavy fermion materials
CeMIn$_5$ (M=Ir, Co, and Rh), with the HoCoGa$_5$-type tetragonal structure.
\cite{Takimoto}

Among CeMIn$_5$ materials, the de Haas-van Alphen (dHvA) effect has been
successfully observed in CeIrIn$_5$ and CeCoIn$_5$,\cite{Haga,Settai}
both of which have huge electronic specific heat coefficients of
several hundreds of mJ/K$^{2}\cdot$mol.
The angular dependence of major experimental dHvA frequency branches is
well explained by a quasi two-dimensional Fermi surface.\cite{Haga,Settai}
This is a clear advantage when we construct a model Hamiltonian,
since it is enough to consider the two-dimensional case.

In order to analyze further the electronic properties from a quantitative
viewpoint, it is desirable to include more precisely
the effect of the actual crystal structure.
For instance, in the previous subsection, we have introduced the level
splitting $\varepsilon$ between $\Gamma_8$ levels in a simple way,
in order to take into account partially the effect of a tetragonal CEF.
However, in an actual tetragonal situation, there occur three Kramers doublets,
two $\Gamma_7$ and one $\Gamma_6$.
Thus, in this subsection, we attempt to incorporate the effect of a tetragonal CEF
more faithfully into the model.

First let us consider the tetragonal CEF term.
After the diagonalization, we obtain three Kramers doublets,
two $\Gamma_7$ and one $\Gamma_6$, expressed as
\begin{equation}
\begin{array}{rcl}
|\Gamma_7^{(1)} \rangle
 &=& (p a^{\dag}_{{\bf i} \pm 5/2} + q a^{\dag}_{{\bf i} \mp 3/2}) |0\rangle, \\
|\Gamma_7^{(2)} \rangle
 &=& (-q a^{\dag}_{{\bf i} \pm 5/2} + p a^{\dag}_{{\bf i} \mp 3/2}) |0\rangle, \\
|\Gamma_6 \rangle
 &=& a^{\dag}_{{\bf i} \pm 1/2} |0\rangle.
\end{array}
\end{equation}
The coefficients $p$ and $q$ are determined by the CEF parameters as
\begin{equation}
p=\sqrt{(1+\alpha)/2},~q = \theta \sqrt{(1-\alpha)/2},
\end{equation}
where $\theta$=$B_4^4/|B_4^4|$ and
\begin{equation}
  \alpha=12(B_2^0+20B_4^0)/\varepsilon_7,
\end{equation}
with
\begin{equation}
  \varepsilon_7=12\sqrt{(B_2^0+20B_4^0)^2+20(B_4^4)^2}.
\end{equation}
The corresponding eigen energies are given by
\begin{equation}
\begin{array}{rcl}
  E(\Gamma_7^{(1)}) &=& 4(B_2^0-15B_4^0)+\varepsilon_7/2, \\
  E(\Gamma_7^{(2)}) &=& 4(B_2^0-15B_4^0)-\varepsilon_7/2, \\
  E(\Gamma_6)&=&-8(B_2^0-15B_4^0),
\end{array}
\end{equation}
respectively.

If the $\Gamma_6$ level is higher than the other two $\Gamma_7$ levels,
it is possible to obtain another simplified model for the $\Gamma_7$ levels.
In fact, in some experimental papers on CeMIn$_5$, we have found that
the position of the $\Gamma_6$ is the highest among the
three Kramers doublets.\cite{Takeuchi,Shishido}
Thus, a simplified and realistic model for CeMIn$_5$
can be constructed by including only two $\Gamma_7$ levels.

\begin{figure}
\includegraphics[width=1.0\linewidth,height=5.6truecm]{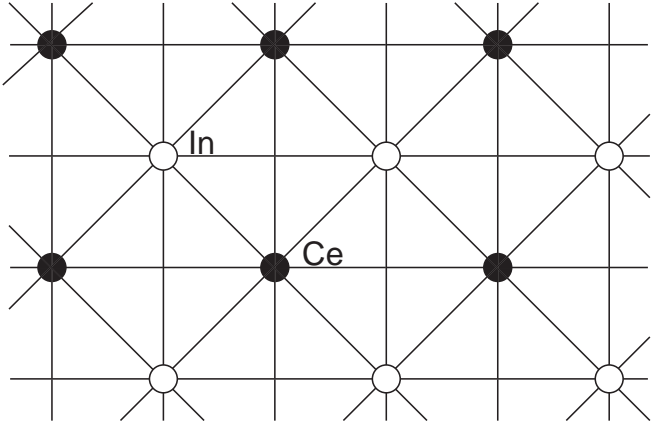}
\caption{
Schematic view of a two-dimensional lattice composed of Ce and In
ions for CeMIn$_5$ (M=Ir, Rh, and Co).
}
\end{figure}

In this crystal structure, the Ce ions form a two-dimensional
tetragonal lattice, while the In ions are located in the
face-centered positions of this Ce-ion network, as shown in Fig.~3.
We include the nearest-neighbor hopping between Ce $4f$ orbitals as well as
next-nearest neighbor hopping via In $5p$ orbitals.
Due to the symmetry, the latter process contributes only along
the diagonal direction,
while hopping via $p$-orbitals along the bond direction vanishes.

Based on the same two-dimensional lattice, we have further
constructed a Hamiltonian called the ``$f$-$p$ model'' by keeping
the $f$-$p$ hybridization explicitly.\cite{Maehira2}
In order to compare the tight-binding dispersion directly with
the results of a relativistic band-structure calculation, it is important
to keep the hybridization term.
However, here we consider an $f$-electron model wherein we include the effects
of $f$-$p$ hybridization in the next-nearest neighbor hopping of $f$ electrons.
We believe it will be adequate for the treatment of superconductivity and magnetism
if we further add the Coulomb interaction terms to the model.

Since two $\Gamma_7$'s are given by the linear combinations of 
$\mu$=$\pm$3/2 and $\pm$5/2, we obtain the $\Gamma_7$ model just by
discarding the $\mu$=$\pm$1/2 states.
It is convenient to define additional new operators as
\begin{equation}
\begin{array}{rcl}
&& c_{{\bf i}1 \uparrow}=a_{5/2},~c_{{\bf i}1 \downarrow}=a_{-5/2}, \\
&& c_{{\bf i}2 \uparrow}=a_{-3/2},~c_{{\bf i}2 \downarrow}=a_{3/2}.
\end{array}
\end{equation}
Again, we can easily show the relation
${\cal K}c_{{\bf i}\tau \sigma}$=$\sigma c_{{\bf i}\tau -\sigma}$.

For the nearest neighbor hopping, we obtain 
\begin{equation}
t_{\tau\tau'}=(t/6)
\left(
\begin{array}{cc}
5 & \sqrt{5}   \\
\sqrt{5} & 1 \\
\end{array}
\right).
\end{equation}
Note that $t$ has the same definition as for the $\Gamma_8$ model.
On the other hand, for the next-nearest neighbor hopping
via the $p$ orbitals, we obtain
\begin{equation}
t'_{\tau\tau'}=(t'/6)
\left(  
\begin{array}{cc}
5 & -\sqrt{5}   \\
-\sqrt{5} & 1 \\
\end{array}
\right),
\end{equation}
for both $[1,1,0]$ and $[1,-1,0]$ directions.
The next-nearest neighbor hopping amplitude is given by
\begin{equation}
t'= {9 \over 28} {(fp\sigma)^2 \over \varepsilon_{\rm f}-\varepsilon_{\rm p}},
\end{equation}
where $(fp\sigma)$ is the Slater-Koster two-center integral for $f$ and $p$ orbitals.
Note that in general $t'$$<$0.

The $\Gamma_7$ model is then written as 
\begin{eqnarray}
H_7 &=&
\sum_{{\bf i,a},\sigma,\tau,\tau'} t_{\tau\tau'}
c^{\dag}_{{\bf i}\tau\sigma}c_{{\bf i+a}\tau'\sigma}
+\sum_{{\bf i,b},\sigma,\tau,\tau'} t'_{\tau\tau'}
c^{\dag}_{{\bf i}\tau\sigma}c_{{\bf i+b}\tau'\sigma} \nonumber \\
&+&(\varepsilon_7/2)\sum_{{\bf i},\sigma,\tau,\tau'}
(\alpha \tau_z + \theta \sqrt{1-\alpha^2} \tau_x)_{\tau \tau'}
c^{\dag}_{{\bf i}\tau\sigma}c_{{\bf i}\tau'\sigma}
\nonumber \\
&+& \sum_{{\bf i},\tau}
U_{\tau}\rho_{{\bf i}\tau\uparrow} \rho_{{\bf i}\tau\downarrow}
+U' \sum_{\bf i} \rho_{{\bf i}a} \rho_{{\bf i}b}
\nonumber \\
&+& J \sum_{{\bf i},\sigma} \rho_{{\bf i}1\sigma} \rho_{{\bf i}2\sigma}
+ J' \sum_{{\bf i},\tau \ne \tau'}
c_{{\bf i}\tau\uparrow}^{\dag} c_{{\bf i}\tau'\downarrow}^{\dag}
c_{{\bf i}\tau\downarrow} c_{{\bf i}\tau'\uparrow},
\end{eqnarray}
where ${\bf a}$ and ${\bf b}$, respectively, denote the vectors connecting 
nearest- and next-nearest neighbor sites.
$\tau_x$ and $\tau_z$ are Pauli matrices.
The energy level shift has been neglected here for simplicity.
The intra-orbital Coulomb interactions are given by
$U_1$=$E_0$$-$$3E_2$+$E_1$ and $U_2$=$E_0$+$5E_2$+$E_1$
for the 1 and 2 orbitals, respectively,
and the inter-orbital Coulomb interaction
is denoted by $U'$=$E_0$$-$$5E_2$.
The Hund's coupling and the pair-hopping interactions are given by
$J$=$10E_2$ and $J'$=$E_1$, respectively.

Three comments on $H_7$ are in order.
(1) There is no transverse component in the Hund's coupling term.
(2) $U_1$$\ne$$U_2$ since the 1 and 2 orbitals belong to different
symmetries.
(3) For the case of $n$=1, both $J$ and $J'$ are sometimes neglected
for simplicity, since they become irrelevant in the large-$U'$ limit.
When we are interested in the strongly correlated region,
it is possible to safely set $E_1$=$E_2$=0 for the case of $n$=1,
leading to $U_1$=$U_2$=$U'$=$E_0$.

%
%
\section{Analysis and Results}

In the previous section, we have discussed in detail the construction of
a microscopic model for $f$-electron systems based on the $j$-$j$ coupling
scheme.
For practical purposes, we have also prepared a couple of simplified
models corresponding to an actual crystal structure.
Those models can be analyzed with several techniques from both
an analytical and a numerical viewpoint.
It is impossible to show here the results for all models.
In this section, we show exact diagonalization results for
$H_8$ as a typical example of an orbital degenerate system.

\subsection{Measured quantities}

Before proceeding to the exhibition of calculated results, let us
discuss the quantities we would measure in a model including orbital degree
of freedom.
First of all, in order to see the spin and charge structure, it is quite
natural to measure the spin and charge correlation functions, defined as
\begin{equation}
  S({\bf q})
  =\sum_{\bf i,j}e^{i{\bf q}\cdot({\bf i-j})}
  \langle s_{z{\bf i}} s_{z{\bf j}} \rangle,
\end{equation}
and 
\begin{equation}
  C({\bf q})
  =\sum_{\bf i,j}e^{i{\bf q}\cdot({\bf i-j})}
  \langle \rho_{{\bf i}} \rho_{{\bf j}} \rangle,
\end{equation}
respectively, 
where
$s_{z{\bf i}}=
\sum_{\tau}(\rho_{{\bf i}\tau\uparrow}$$-$$\rho_{{\bf i}\tau\downarrow})/2$
and 
$\rho_{{\bf i}}=
\sum_{\tau\sigma}\rho_{{\bf i}\tau\sigma}$.
On the other hand, in order to see the orbital structure, it is quite useful
to calculate the orbital correlation function, given by
\begin{equation}
  O_{\tau\tau'}({\bf q})
  =\sum_{\bf i,j}e^{i{\bf q}\cdot({\bf i-j})}
\langle \rho_{{\bf i}\tau} \rho_{{\bf j}\tau'} \rangle.
\end{equation}
As is easily understood from the above equations,
even if the orbital degree of freedom is further included
in the model Hamiltonian,
there is no ambiguity in the definitions for spin and charge correlations,
given by sums in terms of an orbital index.
For the orbital correlations, we also consider as an orbital-dependent
charge correlation, which can be defined without any difficulty.

However, when we attempt to discuss the superconducting pair correlation,
we immediately encounter some serious problems.
Namely, it is necessary to determine the form factors for the pairing states
in both momentum and orbital space.
Note that even in the single-orbital Hubbard model,
it is necessary to determine the momentum dependence,
but in general, our task can be reduced through the help
of a group-theoretical analysis.
In the orbital degenerate model, the extra task of determining the form factor
in orbital space is also imposed upon us.
Especially in the present realistic models,
in which the orbital degree of freedom does $not$ provide a good quantum number,
the situation seems to be very severe at first glance,
since we cannot resort to a group-theoretical analysis
for the orbital dependence of the form factor.

In order to overcome this problem, we carefully reconsider it
based on the concept of off-diagonal long-range order.\cite{Yang}
Let us define the pair correlation as
\begin{equation}
 P^{\pm}=\langle \Phi_{\pm} \Phi_{\pm}^{\dag} \rangle,
\end{equation}
where the subscript $-$ ($+$) denotes spin singlet (triplet)
pairing and $\Phi_{\pm}$ is a pair operator, given in general by
\begin{equation}
 \Phi_{\pm}=\sum_{{\bf i,a},\mu,\mu'}
 \varphi^{\pm}_{{\bf a}\mu\mu'}
 (c_{{\bf i}\mu \uparrow}c_{{\bf i+a}\mu' \downarrow} \pm 
 c_{{\bf i}\mu \downarrow}c_{{\bf i+a}\mu' \uparrow})/\sqrt{2}.
\end{equation}
Since a pair with zero total momentum is considered here,
$\varphi^{\pm}_{{\bf a}\mu\mu'}$ depends only on ${\bf a}$,
which is the vector connecting the sites.
Note also that the orbital dependence of $\varphi^{\pm}_{{\bf a}\mu\mu'}$
is not solely determined by symmetry considerations.

We can easily show that the pair correlation is written in the form
\begin{equation}
 P^{\pm} = \sum_{{\bf a,b},\mu,\nu,\mu',\nu'}
 \varphi^{\pm}_{{\bf a}\mu\mu'}p^{\pm}_{{\bf a}\mu\mu',{\bf b}\nu\nu'}
 \varphi^{\pm}_{{\bf b}\nu\nu'},
\end{equation}
where the matrix element $p^{\pm}_{{\bf a}\mu\mu',{\bf b}\nu\nu'}$
is written as
\begin{eqnarray}
 p^{\pm}_{{\bf a}\mu\mu',{\bf b}\nu\nu'}
 &=&
 \sum_{\bf i,j}
 [\langle 
 c_{{\bf i}\mu \uparrow}
 c_{{\bf i+a}\mu' \downarrow}
 c^{\dag}_{{\bf j+b}\nu' \downarrow}
 c^{\dag}_{{\bf j}\nu \uparrow}
 \rangle \nonumber \\
 &+&
 \langle 
 c_{{\bf i}\mu \downarrow}
 c_{{\bf i+a}\mu' \uparrow}
 c^{\dag}_{{\bf j+b}\nu' \uparrow}
 c^{\dag}_{{\bf j}\nu \downarrow}
 \rangle \nonumber \\
 &\pm&
 \langle 
 c_{{\bf i}\mu \uparrow}c_{{\bf i+a}\mu' \downarrow}
 c^{\dag}_{{\bf j+b}\nu' \uparrow}c^{\dag}_{{\bf j}\nu \downarrow}
 \rangle \nonumber \\
 &\pm&
 \langle 
 c_{{\bf i}\mu \downarrow}c_{{\bf i+a}\mu' \uparrow}
 c^{\dag}_{{\bf j+b}\nu' \downarrow}c^{\dag}_{{\bf j}\nu \uparrow}
 \rangle]/2.
\end{eqnarray}
Here it is necessary to recall the concept of off-diagonal long-range order.\cite{Yang}
Namely, the occurrence of superconductivity is detected when the largest eigenvalue
of the pair correlation matrix becomes the order of $N$, where $N$ is
the number of electrons.
Thus, in a small cluster calculation, the possible superconducting pair
state should be defined by the eigenstate with the maximum eigenvalue
$\rho_{\rm max}$ of the matrix $P$.
Of course, in order to prove the existence of off-diagonal long-range order,
we need to show that $\rho_{\rm max}$ becomes the order of $N$ with increasing
the cluster size. However, at some cluster size, a possible pairing state is
determined without ambiguity by the eigenstate with $\rho_{\rm max}$.

A prescription to define the pair operator is briefly summarized.
First, we determine the pair correlation matrix
$p^{\pm}_{{\bf a}\mu\mu',{\bf b}\nu\nu'}$.
Then we diagonalize it, and the form factor
$\varphi^{\pm}_{{\bf a}\mu\mu'}$ will be defined by
the eigenstate corresponding to the maximum eigenvalue of the
pair correlation matrix.

\subsection{Symmetry of the orbital dependent Cooper-pair}

In the previous subsection, we have discussed a way to define the 
pair operator in orbital degenerate systems.
Here we further consider the symmetry of the Cooper pair in orbital
space from a general point of view.
Anomalous Green's functions for spin triplet and singlet pairs are,
respectively, defined by
\begin{equation}
F^{\rm t}_{\mu\nu}({\bf k},\tau) =
- [\langle c_{{\bf k}\mu\uparrow}(\tau) c_{{\bf -k}\nu\downarrow} \rangle
+ \langle c_{{\bf k}\mu\downarrow}(\tau) c_{{\bf -k}\nu\uparrow} \rangle],
\end{equation}
and
\begin{equation}
F^{\rm s}_{\mu\nu}({\bf k},\tau) =
- [\langle c_{{\bf k}\mu\uparrow}(\tau) c_{{\bf -k}\nu\downarrow} \rangle
- \langle c_{{\bf k}\mu\downarrow}(\tau) c_{{\bf -k}\nu\uparrow} \rangle],
\end{equation}
where 
$c_{{\bf k}\mu\sigma}(\tau)$=$e^{H \tau} c_{{\bf k}\mu\sigma} e^{-H\tau}$ and
$\tau$ denotes imaginary time.
Due to cyclic invariance of the trace, we obtain the following relations:
\begin{equation}
\begin{array}{rcl}
 F^{\rm s}_{\mu\nu}({\bf -k},\tau)
 &=& -F^{\rm s}_{\nu\mu}({\bf k},\beta-\tau),\\
 F^{\rm t}_{\mu\nu}({\bf -k},\tau)
 &=& F^{\rm t}_{\nu\mu}({\bf k},\beta-\tau),
\end{array}
\end{equation}
where $\beta$ is the inverse temperature.
Further, after Fourier transformation, we obtain
\begin{equation}
\begin{array}{rcl}
 F^{\rm s}_{\mu\nu}({\bf -k},i\omega_n)
 &=& F^{\rm s}_{\nu\mu}({\bf k},-i\omega_n),\\
 F^{\rm t}_{\mu\nu}({\bf -k},i\omega_n)
 &=& - F^{\rm t}_{\nu\mu}({\bf k},-i\omega_n),
\end{array}
\end{equation}
where $\omega_n$ is the fermion Matsubara frequency, given by
$\omega_n$=$\pi T$(2$n$+1) with $n$ an integer.
These are the basic relations representing the symmetry of Cooper pairs in
orbital degenerate systems.

Except for differences in the spin states such as singlet and triplet,
``parity'' is the only extra good quantum number to specify the pairing state,
since there is no conservation law regarding orbital degrees of freedom.
Thus, we can produce even- and odd-parity states, satisfying
\begin{equation}
 {\cal P}F^{\rm s}_{\pm}({\bf k},i\omega_n)
 = \pm F^{\rm s}_{\pm}({\bf k},i\omega_n),
\end{equation}
where ${\cal P}$ indicate the operator to change ${\bf k}$ to ${\bf -k}$.
It is easy to understand that $F^{\rm s}_{\pm}$ is given by
linear combinations of $F^{\rm s}_{\mu\nu}({\bf k},i\omega_n)$
and $F^{\rm s}_{\mu\nu}({\bf k},-i\omega_n)$.
Although we cannot define $F^{\rm s}_{\pm}$ uniquely, it is convenient
to consider combinations of odd and even functions
both for orbital exchange and sign-change in frequency.
If we introduce two orbitals ``a'' and ``b'', then
after some simple algebra, we obtain
\begin{equation}
  F^{\rm s}_{\pm}({\bf k},i\omega_n) 
   =\sum_{i=1}^{4} A_{i}^{\pm} F_{i}^{\pm}({\bf k},i\omega_n),
\end{equation}
where $A_{i}^{\pm}$ is an arbitrary number and $F_{i}^{\pm}({\bf k},i\omega_n)$
is given by
\begin{equation}
\begin{array}{rcl}
F_{1}^{\pm}({\bf k},i\omega_n) &=&
F^{\rm s}_{\rm aa}({\bf k},i\omega_n)
\pm F^{\rm s}_{\rm aa}({\bf k},-i\omega_n), \\
F_{2}^{\pm}({\bf k},i\omega_n) &=&
F^{\rm s}_{\rm bb}({\bf k},i\omega_n)
\pm F^{\rm s}_{\rm bb}({\bf k},-i\omega_n), \\
F_{3}^{\pm}({\bf k},i\omega_n) &=&
F^{\rm s}_{\rm ab}({\bf k},i\omega_n)
\pm F^{\rm s}_{\rm ab}({\bf k},-i\omega_n) \\
&+& F^{\rm s}_{\rm ba}({\bf k},i\omega_n)
\pm F^{\rm s}_{\rm ba}({\bf k},-i\omega_n), \\
F_{4}^{\pm}({\bf k},i\omega_n) &=&
F^{\rm s}_{\rm ab}({\bf k},i\omega_n)
\mp F^{\rm s}_{\rm ab}({\bf k},-i\omega_n) \\
&-& F^{\rm s}_{\rm ba}({\bf k},i\omega_n)
\pm F^{\rm s}_{\rm ba}({\bf k},-i\omega_n).
\end{array}
\end{equation}
Since the present Hamiltonian is $not$ invariant with orbital exchange,
\cite{note:symmetry}
the terms ``orbital symmetric'' and ``orbital antisymmetric'' cannot
be used as a label to specify the state.
However, it is convenient to categorize $F_{i}^{\pm}({\bf k},i\omega_n)$
as follows:
The $F_{i}^{+}({\bf k},i\omega_n)$ ($i$=1$\sim$3)
are orbital-symmetric and even function of frequency,
while the $F_{i}^{-}({\bf k},i\omega_n)$ ($i$=1$\sim$3)
are orbital-symmetric and odd function of frequency.
On the other hand, $F_{4}^{+}({\bf k},i\omega_n)$
is orbital-antisymmetric and an odd function of frequency,
while $F_{4}^{-}({\bf k},i\omega_n)$
is orbital-antisymmetric and an even function of frequency.

If we consider the zero-frequency limit, we can simply drop 
the odd functions of frequency.
Namely, the even-parity pair is expressed as a combination
of orbital-symmetric functions, while 
the odd-parity solution is just given by $F_{4}^{-}({\bf k},i\omega_n)$,
the orbital-antisymmetric part.
Note, however, that this is just an accident due to the vanishing of
odd-functions of frequency.
We must $not$ consider that the orbital state provides a good label
to specify the pairing symmetry, although it is a convenient way
to classify the pair state.
Again, it is stressed that spin state and parity are labels used to
specify the pairing symmetry.

For the case of spin triplet, it is easy to repeat the same discussion
as that for spin singlet.
Note that in the zero-frequency limit, the odd-parity solution is given
by the combination of orbital-symmetric parts, while  the even-parity
function is expressed only by the orbital-antisymmetric part.

Up to now we have discussed the general situation in which the system
is $not$ invariant under the exchange of orbitals.
We believe that this is quite natural since in actual materials,
the shapes of orbitals are different from one another.
However, it is possible to consider a fictitious Hamiltonian which is
invariant under orbital exchange.
In such a situation,
the words ``orbital symmetric'' and ``orbital antisymmetric''
recover physical meanings which distinguish the pairing states.
Namely, we can further classify anomalous Green's functions as
\begin{equation}
\begin{array}{rcl}
F^{\rm s}_{+{\rm s}}({\bf k},i\omega_n) 
&=& \sum_{i=1}^3 A_{i}^{+} F_{i}^{+}({\bf k},i\omega_n), \\
F^{\rm s}_{+{\rm a}}({\bf k},i\omega_n) 
&=& F_{4}^{+}({\bf k},i\omega_n), \\
F^{\rm s}_{-{\rm s}}({\bf k},i\omega_n) 
&=& \sum_{i=1}^3 A_{i}^{-} F_{i}^{-}({\bf k},i\omega_n), \\
F^{\rm s}_{-{\rm a}}({\bf k},i\omega_n) 
&=& F_{4}^{-}({\bf k},i\omega_n),
\end{array}
\end{equation}
where the subscripts $+$ and $-$ denotes parity even and odd,
while ``s'' and ``a'' indicate the orbital-symmetric and
orbital antisymmetric, respectively.
In this case, the frequency dependence is automatically classified:
Even-function for ${+{\rm s}}$ and ${-{\rm a}}$ and
odd-function for ${+{\rm a}}$ and ${-{\rm s}}$.

Finally, it is again stressed that when we consider pair correlation
in the static limit, at first glance, orbital symmetric and
antisymmetric parts seem to be separated, but such a separation
is spurious owing to the speciality of the static limit.

\subsection{Numerical Results}

Next we show results obtained using exact diagonalization techniques
for an 8-site tilted cluster.
Since we attempt to make some comparison with CeMIn$_5$,
we consider a two-dimensional lattice.
First we consider the case of quarter filling, i.e., $n$=1.
Note here that $J$ and $J'$ are simply neglected,
since they are considered to be irrelevant in the case of $n$=1.
As shown in Fig.~4(a), for $\varepsilon$$<$0.5,
there is no dominant component in $S({\bf q})$,
indicating a paramagnetic (PM) phase,
while for $\varepsilon$$>$0.5, $S(\pi,\pi)$ becomes abruptly
dominant, strongly suggesting an antiferromagnetic (AF) phase.
This abrupt change is due to level crossing, but it is instructive
to see the charge and orbital correlations.
In Fig.~4(b), we show charge correlation,
exhibiting again an abrupt change around $\varepsilon$$\approx$0.5.
A decrease of charge correlation is the signature of a metal-insulator
transition,
since the charge fluctuations are suppressed in the Mott insulator.
As observed in Fig.~4(c), we can see an orbital disordered (OD) phase
for $\varepsilon$$<$0.5 and a ferro-orbital (FO) phase for $\varepsilon$$>$0.5.
By monitoring the change in spin correlations, we can draw the ground-state
phase diagram for $n$=1, as shown in Fig.~4(d).
We see OD/PM phase for small $U$ and FO/AF phase for large $U$ region.

\begin{figure}
\includegraphics[width=1.0\linewidth,height=10.5truecm]{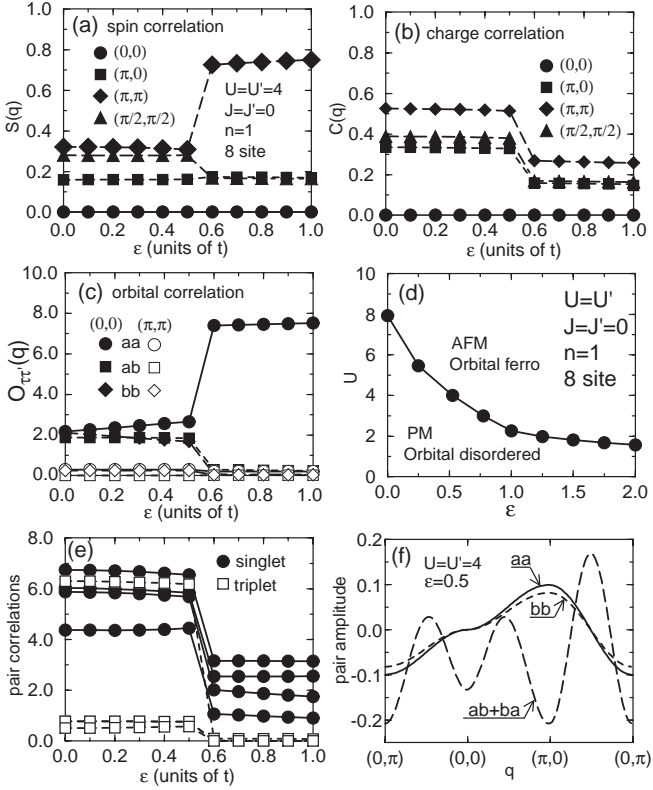}
\caption{
(a) Spin, (b) charge, and (c) orbital correlation functions
vs. $\varepsilon$ for $U$=$U'$=4 and $J$=$J'$=0.
(d) Ground-state phase diagram for the $\Gamma_8$ model for $U$=$U'$
with $J$=$J'$=0.
(e) Pair correlations vs. $\varepsilon$ both for spin singlet and
triplet states.
(f) Pair amplitude in momentum space for the singlet state
having the maximum eigenvalue of the pair correlation matrix.
Note that it is shown separately for $aa$, $bb$,
and $ab$+$ba$ pair states.}
\end{figure}

In Fig.~4(e), we show the eigenvalues of the pair correlation matrices
for spin singlet and triplet states as a function of $\varepsilon$.
The state with the largest eignevalue is a singlet, while a triplet
appears as the second largest eigenvalue state.
Even if we introduce a finite value of $J$, unfortunetely,
this triplet state is $not$ the largest eigen-value state,
at least within the present 8-site calculation.
However, in the previous analysis using the random phase approximation,
the appearance of a triplet state
has been suggested in the small $\varepsilon$ region.\cite{Takimoto}
It may be related to the present result having the triplet state
as the second largest eigenvalue state.

In Fig.~4(f), the pair amplitudes for the largest eigen-value state
is shown in momentum space.
Since we are measuring the static pair correlation,
we can observe here an ``orbital symmetric'' state
with ``aa'', ``bb'', and ``ab+ba'',
while  the ``orbital antisymmetric'' part such as ``ab$-$ba'' is not mixed.
As discussed in the previous subsection, it is just an accident due to
the static limit, and the orbital symmetric state is $not$ 
separated in general.
Next we focus on the momentum dependence of each component.
As seen in Fig.~4(f), the ``aa'' and ``bb'' parts of the pair amplitude
exhibit $B_{\rm 1g}$, while ``ab+ba'' seems to belong to $A_{\rm 1g}$.
If we recall that ``a'' and ``b'' each correspond to $e_{\rm g}$
orbitals, then they have different signs for $\pi/4$ rotation in space.
Thus, there exists a discrepancy in the symmetry between
(``aa'', ``bb'') and ``ab+ba'' components,
although the total symmetry should be $B_{\rm 1g}$ in this case.

We have obtained a simultaneous onset of spin and orbital ordering,
which can be interpreted as a transition between PM and AFM phases
controlled by the suppression of orbital fluctuations.
Moreover, just around the transition regime between PM and AFM
phases, we can expect the appearance of $d$-wave superconductivity
induced by AFM spin fluctuations,
which has been identified as the $B_{\rm 1g}$ pairing state with
the largest eigenvalue.
In fact, the present model has been analyzed using the
random phase approximation, leading to a transition between
PM and $d$-wave superconducting phases with an increase of
$\varepsilon$.\cite{Takimoto}
If we further increase $\varepsilon$, eventually the system becomes
an AFM insulating phase, and this successive transition agrees with
the present results.
Here it is stressed that the orbital degree of freedom plays a key role
to control the ground-state properties. 
We believe that this viewpoint is applicable to the appearance of
$d$-wave superconductivity in CeMIn$_5$,
although the quantitative features should be discussed in detail
using the $\Gamma_7$ model.

\begin{figure}
\includegraphics[width=1.0\linewidth,height=3.5truecm]{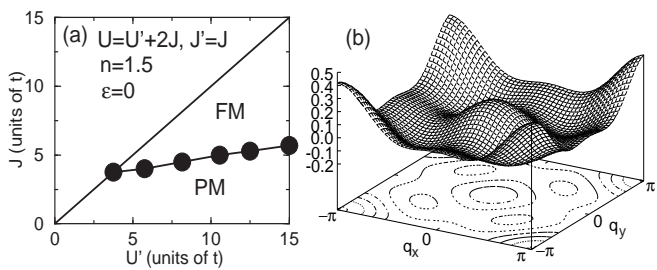}
\caption{
(a) Phase diagram for $\varepsilon$=0 and $n$=1.5.
Note that the region with $U'<J$ is unphysical.
(b) Amplitude for triplet pairing in momentum space, evaluated from
the eigenstate
corresponding to the maximum eigenvalue of the pair correlation matrix.
}
\end{figure}

Next we consider the case of $n$=1.5 for the same $\Gamma_8$ model.
For $\varepsilon$=0, we depict the phase diagram in the ($U'$,$J$) plane,
as shown in Fig.~5(a).
Note that the region with $J$$>$$U'$ is unphysical and here we just ignore it.
We see the appearance of the ferromagnetic (FM) phase, which is stabilized
due to the strong Hund's rule coupling.
The interesting point is the pair amplitude inside the FM phase,
as shown in Fig.~5(b).
The momentum dependence of the largest eigenvalue state is
characterized by even parity.
As naively expected, due to the Hund's rule coupling, a local triplet
is formed, but in order to gain kinetic energy,
this pair should be spatially extended, as shown in Fig.~5(b).
Note here that this is $A_{\rm 1g}$-like, since it is an even-parity state.
Since we are now considering the pairing state in the strong coupling region,
there is no reason to exclude the triplet state with even parity,
which is $not$ directly connected to the usual odd-parity triplet state.
This is not surprising, since local triplet pair must have even parity,
provided that one ion with $f$-orbitals is included in the unit cell.
Thus, quite generally, we can conclude that the odd-parity triplet pair
is not stabilized in the local Hund's rule coupling,
at least in the Bravais lattice, which 
is consistent with the previous discussion.\cite{Norman}

As for the mechanism of the attractive interaction in the pair inside the
FM phase, it should be considered to be orbital fluctuations,
since spin fluctuations are dead in the fully spin-polarized state.
This point is intriguing due to a possible connection with the coexistence
of superconductivity and ferromagnetism in UGe$_2$.
Of course, in order to confirm such an exotic type of superconductivity,
it would be necessary to consider the problem in the bulk limit using,
for instance, the Green's function method.
This case will be discussed elsewhere.

Finally, we note that due care should be paid to the correspondence
between pseudo-spin and total angular momentum ${\bm J}$, 
when we intend to make a comparison with actual materials.
In experiments on $f$-electron compounds, for instance, the external
magnetic field couples to ${\bm J}$, not to the pseudo-spin.
Thus, when we attempt to calculate physical quantities observed
in actuality, it is necessary to consider the response of ${\bm J}$.
The present analysis based on the pseudo-spin model is useful to
see qualitiave tendencies, while a quantiative discussion
is left for future development.

\begin{figure}
\includegraphics[width=1.0\linewidth,height=4.0truecm]{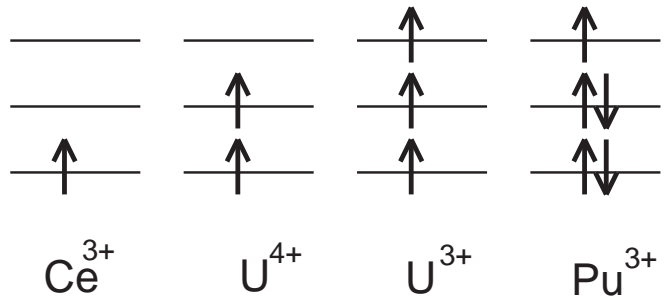}
\caption{
Configuration for $f$-electrons in Ce, U, and Pu ions
for the pseudo spin representations.
Each horizontal bar indicates the pseudo orbital.
}
\end{figure}

%
%
\section{Summary and Discussion}

We have constructed a microscopic model Hamiltonian for $f$-electron
systems based on the $j$-$j$ coupling scheme:
The Coulomb matrix elements have been expressed in terms of the Racah
parameters, and the CEF terms for the two-electron case have been
determined by those for $J$=5/2.
The validity of this $j$-$j$ coupling picture has been carefully
dicsussed in the present paper.
The $f$-electron hopping amplitudes are obtained using
Slater's two-center integral.
Two types of simplified models have also been presented:
One is a Hamiltonian including $\Gamma_8$ and the other
for $\Gamma_7$ levels.
These are believed to be useful for further investigations of
superconductivity and magnetism in $f$-electron systems

In this paper the $\Gamma_8$ model has been analyzed using
the exact diagonalization method.
We have measured several kinds of correlation functions and
developed a prescription to define the pairing state
in multi-orbital systems.
Furthermore, the symmetry of the orbital dependent Cooper pair
has been carefully discussed.
In the analysis of the $\Gamma_8$ model, it has been suggested
that unconventional superconductivity should appear around the
phase boundary between metallic OD-PM and insulating FO-AFM phases.
In addition, we have found evidence for triplet superconductivity 
in the ferromagnetic phase, induced by strong Hund's rule coupling.
With the present method of defining the pairing state,
it has been found that the triplet pairing state is $not$ localized,
but extended spatially to gain kinetic energy.

Based on the $j$-$j$ coupling scheme, it is easy to consider
the case of  several $f$ electrons per site.
Although we have not shown any results in this paper for the case of
$n$=2, corresponding to the $U^{4+}$ ion (see Fig.~6),
we have obtained that the ground-state is an AFM insulator,
which is driven by AFM superexchange coupling
between neighboring $S$=1 spins formed by the local Hund's rule coupling.
When we introduce a level splitting into this AFM insulating state,
we have found robustness for this phase, compared with the
transition around $\varepsilon$=0.5 for the case of $n$=1.
For UMGa$_5$\cite{U115-1,U115-2,U115-3} isostructural with CeMIn$_5$,
if the uranium ions are considered to take U$^{3+}$, again it is naively
deduced to have an AFM ground state, if the local spin is formed
by Hund's rule coupling, as shown in Fig.~6.
However, without any calculations, it is impossible to determine
whether it is a metallic or insulating state, although in experiments,
metallic AFM phase has been suggested.\cite{Kato}
This is left as a future problem.

As introduced in Sec.~1, PuCoGa$_5$ amazingly exhibits superconductivity
with $T_{\rm c}$=18.5K.\cite{PuCoGa5}
In order to understand this high $T_{\rm c}$, first of all,
we generally consider that the spatial extent of the $5f$ wavefunction
is large compared to that of $4f$ electrons,
leading to a larger hybridization among conduction and $5f$ electrons.
Namely, the basic energy scale (or effective bandwidth of the quasiparticles)
in the Pu-compound should be larger than that of Ce-materials.
Here we envision a scenario similar to the high-$T_{\rm c}$ cuprates:
If superconductivity in those ``115'' materials is considered to
originate from an electronic mechanism, $T_{\rm c}$ should be increased
as the energy scale is enhanced.
It may be useful to point out a possible analogy with high-$T_{\rm c}$ materials,
but we emphasize as a separate important issue the active role of orbitals,
which discussed intensively in this paper.
It has been deduced from the local moment behavior
that the plutonium valence is Pu$^{3+}$,\cite{PuCoGa5}
indicating that five $f$-electrons are accommodated.
In our $j$-$j$ coupling picture with pseudo-spin representation,
we can put five electrons in the sextet of $j$=5/2,
as shown in Fig.~6, which is just the hole version of CeMIn$_5$.
Thus, the appearance of superconductivity itself may be understood
without any calculations on the basis of our $j$-$j$ coupling picture.
As explained above, the drastic changes of $T_{\rm c}$ in the ``115''
materials may be understood, based on the difference in the energy scale
for $4f$ and $5f$ electrons, but we believe that another key issue 
related to this problem would be the shape of the orbitals.
A difference between the electron and hole pictures should exist
stemming from the shape of the active orbital, as easily deduced from Fig.~6.
If such differences among orbitals, i.e., level schemes relating to
the CEF effect, will be clarified in those materials,
it may be possible to obtain a unified view to understand unconventional
superconductivity and novel magnetism in ``115'' materials
including CeMIn$_5$, UMGa$_5$, and PuCoGa$_5$.
This point will be further investigated based on the present model.

As repeatedly mentioned in this paper, there remain several open questions
in the electronic properties of $f$-electron materials.
We believe that the microscopic models presented and discussed in this paper
will be useful to advance our understanding of unconventional superconductivity
and novel magnetism in $f$-electron systems.

\section*{Acknowledgement}

The authors thank T. Maehira, K. Miyake, Y. \=Onuki,
T. Takimoto, and K. Yamada for discussions.
We are grateful to R. E. Walstedt for his valuable comments.
The authors are separately supported by the Grant-in-Aid for
Scientific Research from Japan Society for the Promotion of Science.



\begin{references}
\bibitem{note:symmetry}
As mentioned here, the $\Gamma_8$ Hamiltonian $H_8$ is not invariant for
the operation of orbital exchange, but if the operation of spatial
rotational is also combined with orbital exchange,
it is possible to make $H_8$ invariant.
\end{references}
\end{document}